%% file: main.tex
\documentclass[11pt]{article}
\usepackage{algorithm}
\usepackage{algpseudocode}
\usepackage{amsmath}
\usepackage{amssymb}
\usepackage{amsthm}
\usepackage{authblk}
\usepackage{bm}
\usepackage{booktabs}
\usepackage[margin=1in]{geometry}
\usepackage{graphicx}
\usepackage{makecell}
\usepackage{multirow}
\usepackage{tabularx}
\usepackage{xcolor}
\usepackage[colorlinks=true,citecolor=blue]{hyperref}
\usepackage{threeparttable}

\usepackage{cite}
\usepackage[colorlinks=true,citecolor=blue]{hyperref}

\newcommand{\StatexIndent}[2][\algorithmicindent]{%
  \Statex\hspace*{#1}\parbox[t]{\dimexpr\linewidth-#1\relax}{#2}%
}

\newcommand{\LL}{\mathcal{L}}

\newcommand{\GG}{\mathcal{G}}

\graphicspath{{figs/}}

\title{RAMS: Residual-based adversarial-gradient moving sample method for scientific machine learning in solving partial differential equations}

\author[1,2]{Weihang Ouyang}
\author[2]{Min Zhu}
\author[2]{Wei Xiong}
\author[1]{Si-Wei Liu}
\author[2,*]{Lu Lu}
\affil[1]{Department of Civil and Environmental Engineering, Hong Kong Polytechnic University, Hung Hom, Kowloon, Hong Kong, China}
\affil[2]{Department of Statistics and Data Science, Yale University, New Haven, CT, USA}
\affil[*]{Corresponding author. Email: lu.lu@yale.edu}
\date{}

\begin{document}
\maketitle

\begin{abstract}
Physics-informed neural networks (PINNs) and neural operators, two leading scientific machine learning (SciML) paradigms, have emerged as powerful tools for solving partial differential equations (PDEs). Although increasing the training sample size generally enhances network performance, it also increases computational costs for physics-informed or data-driven training. To address this trade-off, different sampling strategies have been developed to sample more points in regions with high PDE residuals. However, existing sampling methods are computationally demanding for high-dimensional problems, such as high-dimensional PDEs or operator learning tasks. Here, we propose a residual-based adversarial-gradient moving sample (RAMS) method, which moves samples according to the adversarial gradient direction to maximize the PDE residual via gradient-based optimization. RAMS can be easily integrated into existing sampling methods. Extensive experiments, ranging from PINN applied to high-dimensional PDEs to physics-informed and data-driven operator learning problems, have been conducted to demonstrate the effectiveness of RAMS. Notably, RAMS represents the first efficient adaptive sampling approach for operator learning, marking a significant advancement in the SciML field.
\end{abstract}


\input{content/introduction}
\input{content/method}
\input{content/examples}
\input{content/conclusion}

\section{Acknowledgments}

This work was partially supported by the Research Grants Council of the Hong Kong Special Administrative Region through the project ``INTACT: Intelligent tropical-storm-resilient system for coastal cities (T22-501/23-R)".

\appendix
\input{content/app_model}
\input{content/app_setup_training_sampling}
\input{content/app_vis}
\input{content/app_abl_proj}

\bibliographystyle{unsrt}
\bibliography{main}

\end{document}

%% file: content/introduction.tex
\section{Introduction}
\label{sec:intro}

Artificial intelligence (AI) has rapidly become an ubiquitous force, revolutionizing various domains within scientific research and engineering applications \cite{jumper2021alphafold, price2025weather, ho2020ddpm}. Among these applications, predicting the solution of partial differential equations (PDEs) using AI is attracting increasing attention from the research community \cite{lin2021operator,yazdani2020systems,brunton2024promising, wang2024artificial,di2023neural,wu2025identifying,wu2025noninvasive, ouyang2025neural}. This burgeoning field leverages machine learning (ML) to address complex scientific challenges governed by the underlying PDEs. Specifically, we aim to train ML models with parameters $\theta$, denoted as $N_{\theta}$, that adhere to constraints imposed by the underlying differential operator $\mathcal{F}$ as $\mathcal{F} \left[ N_{\theta} \right] = 0$.

\begin{figure}[htbp] \label{fig:method_ill}
    \centering
    \includegraphics[width=0.9\linewidth]{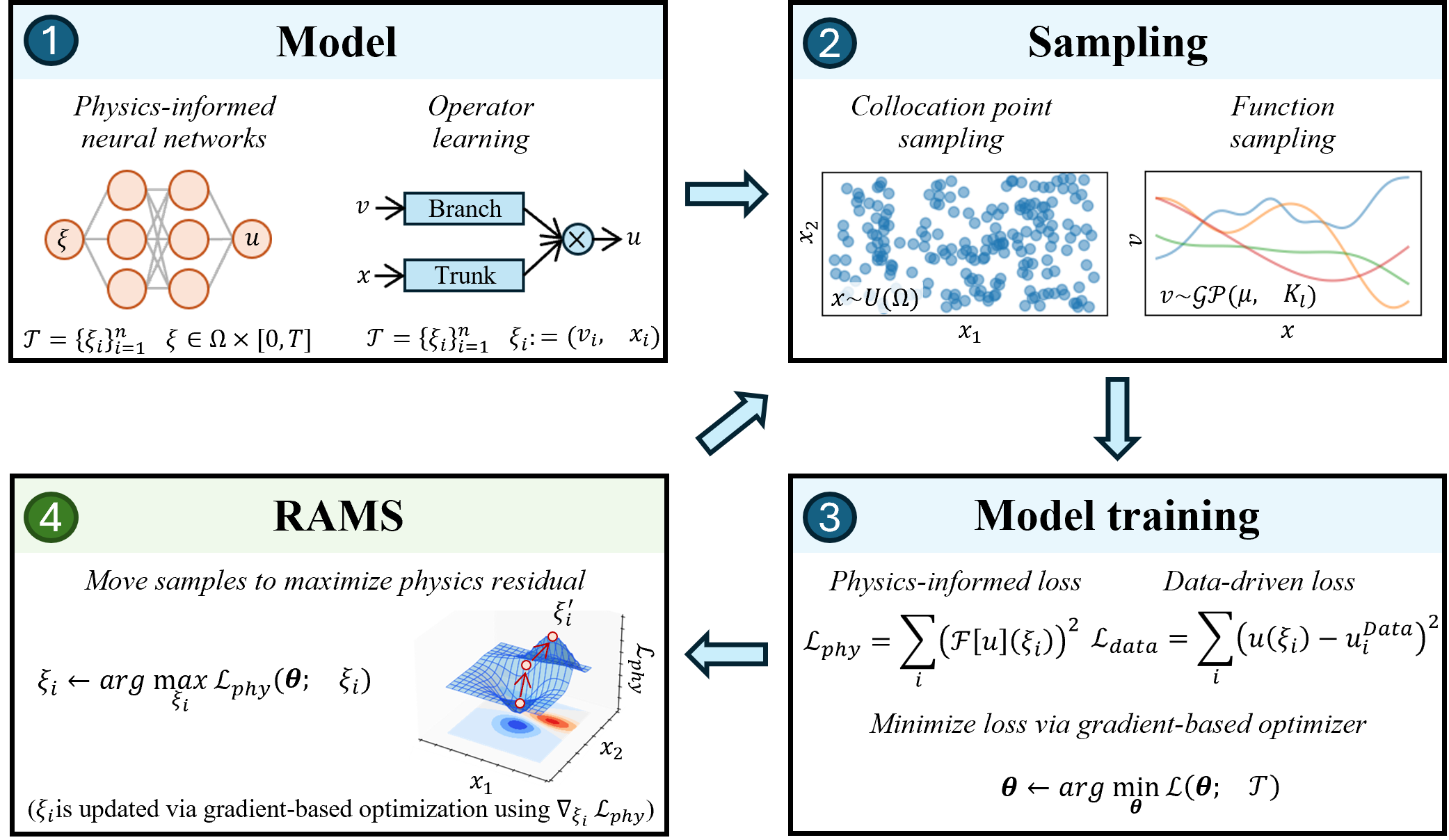}
    \caption{\textbf{Workflow of ML for solving PDEs.} The proposed RAMS treats the samples, either the spatial-temporal coordinates or the functions of interest, as trainable parameters and moves them through the gradient-based optimization to maximize the physics loss defined by the underlying PDEs.}
\end{figure}

Using ML for PDEs typically unfolds in three phases (Fig.~\ref{fig:method_ill}): (1) building a model architecture, (2) collecting training samples, and (3) training the model. To adhere to PDE constraints, the training paradigms of ML models are broadly classified into two categories: data-driven and physics-informed (PI) methods. In the data-driven approach \cite{rudy2019data, wu2020data, dal2020data}, input samples $\mathcal{T}=\{\xi_i\}_i$ and their corresponding outputs $\{u(\xi_i)\}_i$ are generally collected from experimental tests or numerical solvers. The loss function is then formulated based on the discrepancy between the model predictions and the collected data samples. On the other hand, the PI training method \cite{raissi2019physics, lu2021deepxde, karniadakis2021physics, cuomo2022scientific, chen2020physics, zhongkai2024pinnacle} formulates the loss function using the physics residual, $\mathcal{F} \left[ N_{\theta} \right](\xi_i)$, leveraging automatic differentiation~\cite{paszke2017ad} to calculate the derivative terms.
A representative method is the physics-informed neural network (PINN) \cite{raissi2019physics, lu2021deepxde,pang2019fpinns,zhang2019quantifying,lu2021physics}, whose loss is defined by the PDE residual, allowing ML models to predict PDE solutions without data.
Following these two training paradigms, various ML methods have been introduced. The existing conventional ML models, such as the multi-layer perceptron (MLP) and the long-short term memory \cite{zhang2021dive, zhang2020lstm}, have been broadly employed to predict PDE solutions \cite{pinkus1999approximation}. 
However, these models may underperform in some complex scenarios, such as operator learning, where the models are required to approximate mappings between function spaces \cite{lu2021dno, xiong2024koopman}. To address this, Lu et al. \cite{lu2021dno} introduced a new ML model, the deep operator network (DeepONet), based on the universal approximation theory for operators \cite{chen1993approximations}. After that, extensive operator learning models have been proposed \cite{lu2022comprehensive_no, lu2022multifidelity, zhu2023fourier, yin2024scalable, cai2021deepm,mao2021deepm,bausback2025stochastic,wang2025fundiff,xiao2025quantum}, e.g., MIONet \cite{jin2022mionet} for multiple inputs, Fourier neural operator \cite{li2020fno,lee2024efficient}, Wavelet neural operator \cite{tripura2023wno}, and Laplace neural operator \cite{cao2024lno}.

Both data-driven and PI training paradigms are fundamentally reliant on training samples, for either collecting data samples or computing physics residuals at sample points. These training paradigms inevitably face a trade-off between the number of samples and model performance. Utilizing more samples increases computational demands for data collection and model training processes, while insufficient samples might deteriorate model performance \cite{nabian2021efficient}. In response, some studies have been conducted on developing efficient sampling methods that use fewer samples without compromising performance. A notable contribution is the adaptive sampling method by residual-based adaptive refinement (RAR)~\cite{lu2021deepxde,yu2022gradient}. This technique strategically selects collocation points in areas exhibiting large physics residuals, which are indicative of regions that are insufficiently trained. Following RAR, similar adaptive sampling strategies, such as RAR-D \cite{wu2023comprehensive, hanna2022rard} and R3 \cite{daw2022r3}, have been introduced to target collocation points around areas with significant physics residuals. Beyond these RAR-type methods, other sampling strategies for PINNs have been developed using importance sampling (IS) \cite{tokdar2010is} to minimize the generalization error, where collocation points are sampled according to the probability distribution aligned with the physics residual \cite{nabian2021efficient, tang2023das, wang2024das2, zhang2025annealed}. Despite the intent to reduce the variance in estimating physics residual norms, these IS-based methods share a similar feature with RAR-type methods as they are all designed to densely sample in regions with large residuals \cite{daw2022r3}.

Even though these methods with dense sampling around areas with large physics residuals have been proven effective in many PINN examples, they are still less capable for some scenarios, such as high-dimensional PDE problems. In these cases, effectively sampling around areas with large physics residuals becomes computationally daunting due to the sparsity of the high-dimensional space. Furthermore, given that operator learning commonly requires parametrizing the input functions as high-dimensional vectors, no feasible adaptive sampling strategy for operator learning has been proposed. For instance, Wang et al. \cite{wang2024das2} proposed a novel sampling method, DAS\textsuperscript{2}, using a flow model to estimate the physics residual distribution as the underlying sampling distribution for high-dimensional problems. However, applying DAS\textsuperscript{2} to operator learning problems requires a prohibitively large number of function samples to yield reliable results. This underscores the urgency for the development of efficient sampling methods, particularly in high-dimensional contexts.

To address this challenge, we introduce a new gradient-based optimization method, residual-based adversarial-gradient moving sample (RAMS), to move samples according to the adversarial gradient for maximizing the physics residuals (Fig.~\ref{fig:method_ill}). RAMS is applicable for both PINN and operator learning methods and can be conveniently integrated with any sampling methods. RAMS can also be used for data-driven problems and significantly improve model performance. Extensive numerical examples have yielded several key observations.
\begin{itemize} 
\item RAMS exhibits great compatibility with existing sampling approaches for PINNs, leading to orders of magnitude improvement in accuracy without increasing the number of collocation points. 
\item RAMS substantially reduces the computational cost when applied to solving high-dimensional problems. 
\item RAMS has consistent improvement in accuracy for both PI and data-driven operator learning problems with the same number of training samples. 
\end{itemize}

The paper is organized as follows. In Section~\ref{sec:methods}, we briefly introduce PINN, operator learning, and several classical sampling techniques, based on which the proposed RAMS method is presented. In Section~\ref{sec:ex}, comprehensive tests are conducted across a wide range of problems, including PINN, PI operator learning, and data-driven operator learning, to demonstrate the effectiveness of RAMS. Finally, Section~\ref{sec:con} concludes the paper and outlines potential future research directions.

%% file: content/method.tex
\section{Methods}
\label{sec:methods}

In this section, we first briefly introduce PINN to predict PDE solutions by minimizing the loss function of physics residual. Then, the operator learning using DeepONet is presented for both PI and data-driven training. 
Several sampling methods in PINN are then discussed, based on which our residual-based adversarial-gradient moving sample (RAMS) method is developed.

\subsection{Physics-informed neural networks}

Consider the following PDE with the solution $u$:
\begin{equation*}
    \begin{aligned}
        \mathcal{F}[u](x, t) = 0,& \quad \text{for } (x, t) \in \Omega \times [0,T],\\
    \mathcal{B}[u](x, t) = 0,& \quad \text{for } (x, t) \in \partial \Omega \times [0,T],\\
    u(x, 0) = u_0(x),& \quad \text{for } x \in \Omega,
    \end{aligned}
\end{equation*}
where $\mathcal{F}$ and $\mathcal{B}$ are the differentiation and boundary operators of the PDE, and $u_0$ denotes the initial condition.

In PINN \cite{raissi2019physics}, a neural network $N_{\theta}(\xi)$ is trained to predict the PDE solution, where $\xi := (x, t) \in \Omega \times [0, T]$ is the spatial-temperate coordinate of the PDE solution. In this study, the widely-used multi-layer perceptron (MLP) is employed, which is parametrized as
\begin{equation*}
    \begin{aligned}
        N_{\theta}(\xi) &= g^n \circ \sigma \circ g^{n-1} \circ \sigma \circ \cdots \sigma \circ g^1(\xi), \\
        g^i(\xi) &= W_i \xi + b_i, \quad
        \theta = \left\{ W_i, b_i\right\}_{i=1}^n,
    \end{aligned}
\end{equation*}
where $\sigma$ is a nonlinear activation function, e.g., the hyperbolic tangent function in this work; $n$ denotes the depth of the MLP; and $\theta$ is the collection of trainable parameters of $N_{\theta}$ that contains all the weight matrices $W_i$ and bias vectors $b_i$.

The PINN loss is defined based on the collocation points $\mathcal{T}=\mathcal{T}_{phy} \cup \mathcal{T}_{BC} \cup \mathcal{T}_{IC}$ as
\begin{equation}
    \label{eq:dis_pinn_loss}
    \begin{aligned}
        \LL(\theta;\mathcal{T}) &= \LL_{phy}(\theta;\mathcal{T}) + \LL_{BC}(\theta;\mathcal{T}) + \LL_{IC}(\theta;\mathcal{T}), \\
        \LL_{phy}(\theta;\mathcal{T}) &= \frac{1}{N_{phy}} \sum_{i=1}^{N_{phy}} \left( \mathcal{F}[N_{\theta}](\xi_i^{phy}) \right)^2, \quad
        \mathcal{T}_{phy}=\{\xi_i^{phy} \in \Omega \times [0,T]\}_{i=1}^{N_{phy}}, \\
        \LL_{BC}(\theta;\mathcal{T}) &= \frac{1}{N_{BC}} \sum_{i=1}^{N_{BC}} \left( \mathcal{B}[N_{\theta}](\xi_i^{BC}) \right)^2, \quad 
        \mathcal{T}_{BC}=\{\xi_i^{BC} \in \partial\Omega \times [0,T]\}_{i=1}^{N_{BC}}, \\
        \LL_{IC}(\theta;\mathcal{T}) &= \frac{1}{N_{IC}} \sum_{i=1}^{N_{IC}} \left( N_{\theta}(\xi_i^{IC}) - u_0(\xi_i^{IC}) \right)^2, \quad
        \mathcal{T}_{IC}=\{\xi_i^{IC} \in \Omega \times [0]\}_{i=1}^{N_{IC}},
    \end{aligned}
\end{equation}
where $N$ is the number of the sampled collocation points. The optimal MLP parameters $\theta^*$ are then computed by
\begin{equation*}
    \theta^* = \arg \min_{\theta} \LL(\theta;\mathcal{T}),
\end{equation*}
which is commonly solved via a stochastic gradient descent method, e.g., Adam \cite{kingma2014adam}.

\subsection{Operator learning}

PINN is developed for predicting solutions of a single PDE problem, limiting its broader application. Hence, extensive studies have been conducted to include PDE parameters in the model input $\xi$, enabling PINN to predict parametric PDE solutions after training \cite{gao2021phygeonet, ouyang2024machine, ouyang2024physics}. However, when considering more complicated operator learning problems, PINN is less effective. MLP and other conventional ML models are generally designed to approximate a function, a mapping between two finite-dimensional spaces. However, in operator learning problems, the model should approximate a mapping between two infinite-dimensional Banach spaces, i.e., an operator $\GG$ mapping from a function $v$ to another function $u$ as
\begin{equation*}
    \GG: v \mapsto u,
\end{equation*}
where $v$ could be a function parametrizing the PDE problem, such as the initial condition $u_0$, and $u$ could be the PDE solution.

Unlike conventional ML models, DeepONet \cite{lu2021dno} was developed to address operator learning problems. DeepONet approximates the operator using two networks, a branch net and a trunk net. The branch net takes a finite-dimensional representation of the input function $\Tilde{v}$ as the input (e.g., the function values at certain locations), and the trunk net takes a spatial-temporal coordinate of the output function $(x,t) \in \Omega \times [0,T]$ as the input. Then, the value of the output function at the corresponding coordinate is computed via the element-wise product of the output vectors of the two subnets as
\begin{equation*}
    \GG_{\theta}[v](x,t) = \sum_{i=1}^n N^b_i(\Tilde{v}) N^t_i(x,t) + N^b_0,
\end{equation*}
where $\GG_{\theta}$ is a DeepONet model parameterized by $\theta$; $\left \{N^b_i \right \}_{i=1}^n$ and $\left \{N^t_i \right \}_{i=1}^n$ are the output features from the branch net and the trunk net, respectively; and $N^b_0 \in \mathbb{R}$ is a trainable bias.

DeepONet can be trained using data-driven methods, whose loss function is formulated based on pre-collected data samples. The data-driven loss function is
\begin{equation*}
    \label{eq:dis_ddol_loss}
    \LL_{data}(\theta) = \frac{1}{N_{data}} \sum_{i=1}^{N_{data}} \left( \GG_{\theta}[\Tilde{v}_i](x_i, t_i) - u_i \right)^2,
\end{equation*}
where $u_i$ is the solution data of the input samples including both the function sample $v_i$ and the spatial-temporal collocation point $(x_i, t_i)$.

Similarly, the PI loss in Eq.~\eqref{eq:dis_pinn_loss} can also be applied to DeepONet by sampling both input functions $v$ and collocation points. For instance, training a DeepONet to approximate the mapping from the initial condition to the PDE solution, $\GG: u_0 \mapsto u$, the PI loss for the operator learning is
\begin{equation*}
    \label{eq:dis_piol_loss}
    \LL_{phy}(\theta;\mathcal{T}) = \frac{1}{N_{phy}} \sum_{i=1}^{N_{phy}} \left( \mathcal{F}[\GG_{\theta}[\Tilde{v}^{phy}_i]](x_i^{phy}, t_i^{phy}) \right)^2, \quad
        \mathcal{T}_{phy} = \{ \xi_i^{phy}\}_{i=1}^{N_{phy}},
\end{equation*}
where $\xi_i^{phy} = (\Tilde{v}_i^{phy}, x_i^{phy}, t_i^{phy})$ represents a sample including both the input function and coordinates. The BC loss $\LL_{BC}$ and the IC loss $\LL_{IC}$ are computed similarly to those PINN losses in Eq.~\eqref{eq:dis_pinn_loss}.

\subsection{Sampling methods}

Despite extensive research in SciML, the development of sampling methods for $\mathcal{T}$ mainly focuses on PINN and remains relatively limited for operator learning. A commonly employed technique is random sampling, where collocation points are uniformly randomly sampled over the domain. Beyond uniform sampling, quasi-random low-discrepancy sequences like Latin hypercube sampling (LHS) and Halton sequence sampling are also utilized to reduce sample variance, thereby leading to improved accuracy \cite{wu2023comprehensive}.

Apart from non-adaptive random and quasi-random techniques, adaptive sampling methods have been developed for PINN. The first adaptive sampling method is the residual-based adaptive refinement (RAR) method \cite{lu2021deepxde}, which initially samples some collocation points and then, at each resampling stage, adds more points with the largest PDE residuals into the existing collocation points. Then, a RAR with distribution (RAR-D) sampling method \cite{wu2023comprehensive, hanna2022rard} is further developed, where the resampled collocation points follow the distribution of the PDE residuals. To better distinguish these two methods, the greedy RAR method in Ref.~\cite{lu2021deepxde} is referred to as RAR with greed (RAR-G) in the rest of this paper. Daw et al. \cite{daw2022r3} also proposed a Retain-Resample-Release (R3) sampling method, where the resampled collocation points are selected in a pattern similar to the evolutionary algorithm to aggregate them around the area with large PDE residuals. For a comprehensive comparison of different sampling methods for PINN, we refer the reader to Ref.~\cite{wu2023comprehensive}.

In summary, these adaptive sampling methods typically update the initially sampled collocation points after certain training epochs (Algorithm~\ref{alg:resam_process}). Most of these approaches are developed based on the IS, where new collocation points are sampled according to a specially designed distribution, e.g., the PDE residual. 
These IS-based adaptive sampling methods usually encourage the resampled points congregated around areas with large PDE residuals \cite{daw2022r3, zhang2025annealed}.

\begin{algorithm}
\caption{\textbf{Adaptive sampling for PINN.}}
\begin{algorithmic}[1]
\State Initialize the model $N_{\theta}$;
\State Generate the initial samples $\mathcal{T}$ using a predefined sampling method;
\Repeat
    \State Train the model for a certain number of iterations;
    \State Update the samples $\mathcal{T}$ using the selected sampling method;
\Until{the stop criterion is achieved.}
\end{algorithmic}
\label{alg:resam_process}
\end{algorithm} 

\subsection{Residual-based adversarial-gradient moving sample method}

\subsubsection{Method overview}
\label{sec:rams_overview}

Although existing sampling methods demonstrate promising performance in various PINN applications, the study of high-dimensional PDEs and operator learning problems is still limited, primarily because locating areas with large PDE residuals in high-dimensional spaces is computationally demanding. To address this issue, we develop a residual-based adversarial-gradient moving sample (RAMS) method by leveraging gradient-based optimization techniques. RAMS effectively congregates samples around regions with large PDE residuals by moving samples according to the adversarial gradient direction.

RAMS can be seamlessly integrated with existing sampling techniques by maximizing the PDE residual for the generated samples. In RAMS (Algorithm~\ref{alg:rams}), we treat the samples (collocation points in PINN or function samples in DeepONet) as trainable parameters. These trainable samples are optimized to maximize $\LL_{phy}$ via gradient ascent based on the gradient $\nabla_{\xi} (\mathcal{F} \left[ N_{\theta} \right](\xi))^2$. After optimization, when necessary, the trainable samples are projected back to the original domain through an appropriate projection technique $\mathcal{P}$. For PINN, if the trained samples are outside the domain, a simple projection method is to map the optimized collocation points to the nearest point on the domain boundary. For operator learning, when input functions are generated from a Gaussian random field (GRF) and parameterized using sensor values, we use the kernel smoothing method \cite{wand1994kernel} to smooth the trained samples after each training iteration. This method utilizes the same covariance kernel function $k$ as defined in the GRF to preserve similar function continuity and smoothness. We illustrate the necessity of the kernel smoothing method as the projector in RAMS for operator learning in Section~\ref{sec:piol_diff}.

\begin{algorithm}[htbp]
\caption{\textbf{Residual-based adversarial-gradient moving sample (RAMS) method.} $\hat{\mathcal{T}}$ is the set of all trainable samples determined by the selected sampling method. The updates for all $\xi \in \hat{\mathcal{T}}$ can be vectorized for parallel execution.}
\label{alg:rams}
\begin{algorithmic}[1]
    \For{$\xi \in \hat{\mathcal{T}}$}
    \State Compute the gradient $\nabla_{\xi} \LL_{phy}$ using automatic differentiation;
    \State Update $\xi$ with gradient ascent to maximize $\LL_{phy}$ for $n_{RAMS}$ iterations;
    \State If needed, project $\xi$ back to the sampling space by $\mathcal{P}(\xi)$.
    \EndFor
\end{algorithmic}
\end{algorithm}

To illustrate the kernel smoothing method, we consider a function defined as
\begin{equation*}
    f : \Omega \ni x \mapsto y \in \mathbb{R},
\end{equation*}
where $\Omega$ is the computational domain. The function is parameterized at the sensor locations by
\begin{equation*}
    \bm{f} = [f(x_1), f(x_2), \dots, f(x_n)]^T \in \mathbb{R}^n,
\end{equation*}
where $\{x_i \in \Omega \}_{i=1}^n$ is a set of sensor locations. The projection $\mathcal{P}$ is given by
\begin{equation*}
    \mathcal{P}(\bm{f}) = \bm{K} \bm{f} \oslash \bm{k}',
\end{equation*}
where $\bm{K} = [K_{ij}] \in \mathbb{R}^{n \times n}$ represents the kernel matrix with $K_{ij} = k(x_i, x_j)$ computed via the covariance kernel function; $\oslash$ denotes element-wise division; and $\bm{k}' \in \mathbb{R}^n$ is a vector defined as
\begin{equation*}
    \bm{k}' = \left[\sum_{j=1}^n K_{1j}, \sum_{j=1}^n K_{2j}, \dots, \sum_{j=1}^n K_{nj}\right]^T.
\end{equation*}

In addition to PINN and physics-informed operator learning, we also apply RAMS to data-driven operator learning by guiding the collection of new data, which works as an active learning strategy \cite{kou2025active,winovich2025active}. Specifically, we first collect a small training dataset and then progressively add new data points to the dataset. During each sampling stage, newly generated input function samples are optimized to maximize the PDE residual through RAMS. These new input samples are then processed by a numerical solver or experiment to generate the corresponding ground-truth outputs to form new data points, which are added to the training dataset.

\subsubsection{RAMS for PINN}
\label{sec:RAMS_for_PINN}

In this section, we choose several sampling methods to demonstrate how to integrate RAMS for PINN. First, we apply RAMS to enhance three non-adaptive sampling methods (Algorithm~\ref{alg:trainable_sampling}), including uniformly random sampling (Random), LHS, and Halton sequence~\cite{wu2023comprehensive}. For these non-adaptive sampling methods, we first generate a set of samples and split it into two subsets before network training: a fixed set that remains unchanged and a trainable set. At each resampling stage, we randomly pick a subset from the trainable set and update it by RAMS, while keeping the fixed set unchanged.

\begin{algorithm}
\caption{\textbf{Non-adaptive sampling method (e.g., Random, LHS, and Halton) with RAMS for physics-informed machine learning}.}
\begin{algorithmic}[1]
\State Initialize the network $N_{\theta}$;
\State Train the network for a certain number of iterations using $\mathcal{T}$;
\State Partition $\mathcal{T}$ into a fixed set $\mathcal{T}_1$ and a trainable set $\mathcal{T}_2$;
\For{$i \gets 1$ \textbf{to} $t_r$}
    \State Train the network for $n_{train}$ iterations using $\mathcal{T}$;
    \State Randomly choose a subset $\hat{\mathcal{T}} \subset \mathcal{T}_2$, and denote the remainder as $\mathcal{T}_2'$;
    \State Update $\hat{\mathcal{T}}$ using \textbf{RAMS};
    \State Update the trainable set: $\mathcal{T}_2 \gets \hat{\mathcal{T}} \cup \mathcal{T}_2'$;
    \State Update all samples: $\mathcal{T} \gets \mathcal{T}_1 \cup \mathcal{T}_2$;
  \EndFor
\State Train the network for a certain number of iterations using $\mathcal{T}$.
\end{algorithmic}
\label{alg:trainable_sampling}
\end{algorithm}

We then describe the integration of RAMS with three adaptive sampling methods: RAR-G~\cite{lu2021deepxde,wu2023comprehensive}, RAR-D~\cite{wu2023comprehensive}, and R3~\cite{daw2022r3}. In RAR (including both RAR-G and RAR-D) with RAMS, new samples added at each resampling stage are treated as trainable and updated by RAMS (Algorithm~\ref{alg:rar_with_rams}). In contrast, R3 does not add points near high-residual regions; instead, it retains the points with the largest residuals at each stage. In R3 with RAMS (Algorithm~\ref{alg:r3_with_rams}), these retained points are trainable and updated by RAMS.

\begin{algorithm}
\caption{\textbf{RAR (RAR-G or RAR-D) with RAMS for physics-informed machine learning}.}
\begin{algorithmic}[1]
\State Initialize the network $N_{\theta}$;
\State Sample the initial samples $\mathcal{T}$ of size $n_{ini}$ using the selected sampling method;
\For{$i \gets 1$ \textbf{to} $t_r$}
    \State Train the network for $n_{train}$ iterations using $\mathcal{T}$;
    \State Randomly generate $M$ samples to form $\mathcal{T}_0$;
    \State Select $m$ samples from $\mathcal{T}_0$ to form $\hat{\mathcal{T}}$ using one of:
    \StatexIndent{\textbf{RAR-G}: Choose the $m$ samples with the largest PDE residuals in $\mathcal{T}_0$;}
    \StatexIndent{\textbf{RAR-D}: Sample $m$ points from $\mathcal{T}_0$ according to the probability density function (PDF) proportional to the PDE residual as $p(\xi) \propto \left( \mathcal{F}[N_{\theta}](\xi) \right)^2$;}
    \State Update $\hat{\mathcal{T}}$ using \textbf{RAMS};
    \State Update all samples: $\mathcal{T} \gets \mathcal{T} \cup \hat{\mathcal{T}}$;
\EndFor
\State Train the network for a certain number of iterations using $\mathcal{T}$.
\end{algorithmic}
\label{alg:rar_with_rams}
\end{algorithm}

\begin{algorithm}
\caption{\textbf{R3 with RAMS for physics-informed machine learning}.}
\begin{algorithmic}[1]
\State Initialize the network $N_{\theta}$;
\State Sample the initial samples $\mathcal{T}$ using the selected sampling method;
\For{$i \gets 1$ \textbf{to} $t_r$}
    \State Train the network for $n_{train}$ iterations using $\mathcal{T}$;
    \State Compute the PDE residuals for the samples in $\mathcal{T}$;
    \State Compute the threshold $\tau$ as the average PDE residual;
    \State Select the samples $\hat{\mathcal{T}}$ with the PDE residuals larger than $\tau$;
    \State Update $\hat{\mathcal{T}}$ using \textbf{RAMS};
    \State Randomly generate a new set of samples $\mathcal{T}_s$ of size $|\mathcal{T}| - |\hat{\mathcal{T}}|$;
    \State Update all samples: $\mathcal{T} \gets \mathcal{T}_s \cup \hat{\mathcal{T}}$;
\EndFor
\State Train the network for a certain number of iterations using $\mathcal{T}$.
\end{algorithmic}
\label{alg:r3_with_rams}
\end{algorithm}

\subsubsection{RAMS for operator learning}

Similar to PINN, we also apply RAMS to physics-informed operator learning with random sampling and RAR-G. The workflow follows Algorithms~\ref{alg:trainable_sampling} and \ref{alg:rar_with_rams}. The only difference is that in operator learning, we use GRFs to randomly sample input functions parametrized by sensor values. Moreover, the kernel smoothing method (Section~\ref{sec:rams_overview}) is utilized as the projector in the RAMS method.

In addition, we develop RAR-G with RAMS for data-driven operator learning (Algorithm~\ref{alg:rar_with_rams_ddol}). In this case, we only generate a small training dataset prior to network training, and the network is first trained on this initial dataset. To improve the network accuracy, we randomly generate a large pool of candidate input functions and select the subset with the largest PDE residuals, which are then refined via RAMS. We compute their corresponding outputs and add the resulting input-output pairs to the training dataset for further training.

\begin{algorithm}
\caption{\textbf{RAR-G with RAMS for data-driven operator learning.}}
\begin{algorithmic}[1]
\State Initialize the network $N_{\theta}$;
\State Collect $n_{ini}$ input functions with the corresponding outputs to form the training set $\mathcal{T}$;
\For{$i \gets 1$ \textbf{to} $t_r$}
    \State Train $N_\theta$ using $\mathcal{T}$;
    \State Randomly generate $M$ input functions to form $\mathcal{V}$;
    \State Select the $m$ input functions from $\mathcal{V}$ with the largest PDE residuals to form $\hat{\mathcal{V}}$;
    \State Update $\hat{\mathcal{V}}$ using \textbf{RAMS};
    \State Compute the outputs for $\hat{\mathcal{V}}$ to build a new training dataset $\hat{\mathcal{T}}$.
    \State Update the training dataset: $\mathcal{T} \gets \mathcal{T} \cup \hat{\mathcal{T}}$;
\EndFor
\State Train the network using $\mathcal{T}$.
\end{algorithmic}
\label{alg:rar_with_rams_ddol}
\end{algorithm}

%% file: content/examples.tex
\section{Results}
\label{sec:ex}

In this section, we present ten examples to demonstrate the effectiveness of the proposed RAMS method in PINN (Section~\ref{sec:ex_pinn}), PI operator learning (Section~\ref{sec:ex_piol}), and data-driven operator learning (Section~\ref{sec:ex_ddol}). These examples show that RAMS significantly improves model performance by effectively moving data points to critical areas, all while using the same amount of training data. The network architectures used in all examples are summarized in Appendix~\ref{apd:models}. The sampling and training setups for all examples are summarized in Appendix~\ref{apd:setup}, unless otherwise stated.

\subsection{Physics-informed neural networks}
\label{sec:ex_pinn}

We first investigate the performance of RAMS for PINN and assess its compatibility with various established sampling methods. Specifically, we implement RAMS alongside three non-adaptive sampling methods (uniformly random sampling, LHS, and Halton sequence)~\cite{wu2023comprehensive} and three adaptive sampling methods (RAR-G~\cite{lu2021deepxde}, RAR-D~\cite{wu2023comprehensive}, and R3~\cite{daw2022r3}). We refer to the original versions of these methods as ``w/o RAMS" and the versions augmented by RAMS as ``with RAMS". For each method, we evaluate the performance via the mean relative $L^2$ error from ten independent experiments on three PDE problems, including 1D Burgers' equation (Section~\ref{sec:pinn_burgers}), 1D wave equation (Section~\ref{sec:pinn_wave}), and 2D Poisson equation (Section~\ref{sec:pinn_poisson}). Then, a high-dimensional PDE (up to 10 dimensions; Section~\ref{sec:pinn_hd}) is tested to illustrate the performance of the proposed RAMS in handling high-dimensional problems.

\subsubsection{1D Burgers' equation}
\label{sec:pinn_burgers}

A Burgers' equation with zero Dirichlet boundary conditions is considered here as
\begin{equation*}
    \begin{aligned}
        \frac{\partial u}{\partial t} + u \frac{\partial u}{\partial x} = \nu \frac{\partial^2 u}{\partial x^2}, &\quad \text{for } x \in [-1,1], t \in [0,1], \\
        u(x,0) = -\sin(\pi x), &\quad u(-1,t) = u(1,t) = 0,
    \end{aligned}
\end{equation*}
where $u$ represents the flow velocity and $\nu=\frac{0.01}{\pi}$ is the viscosity of the fluid.

\begin{figure}[htbp] 
    \centering
    \includegraphics[width=0.8\linewidth]{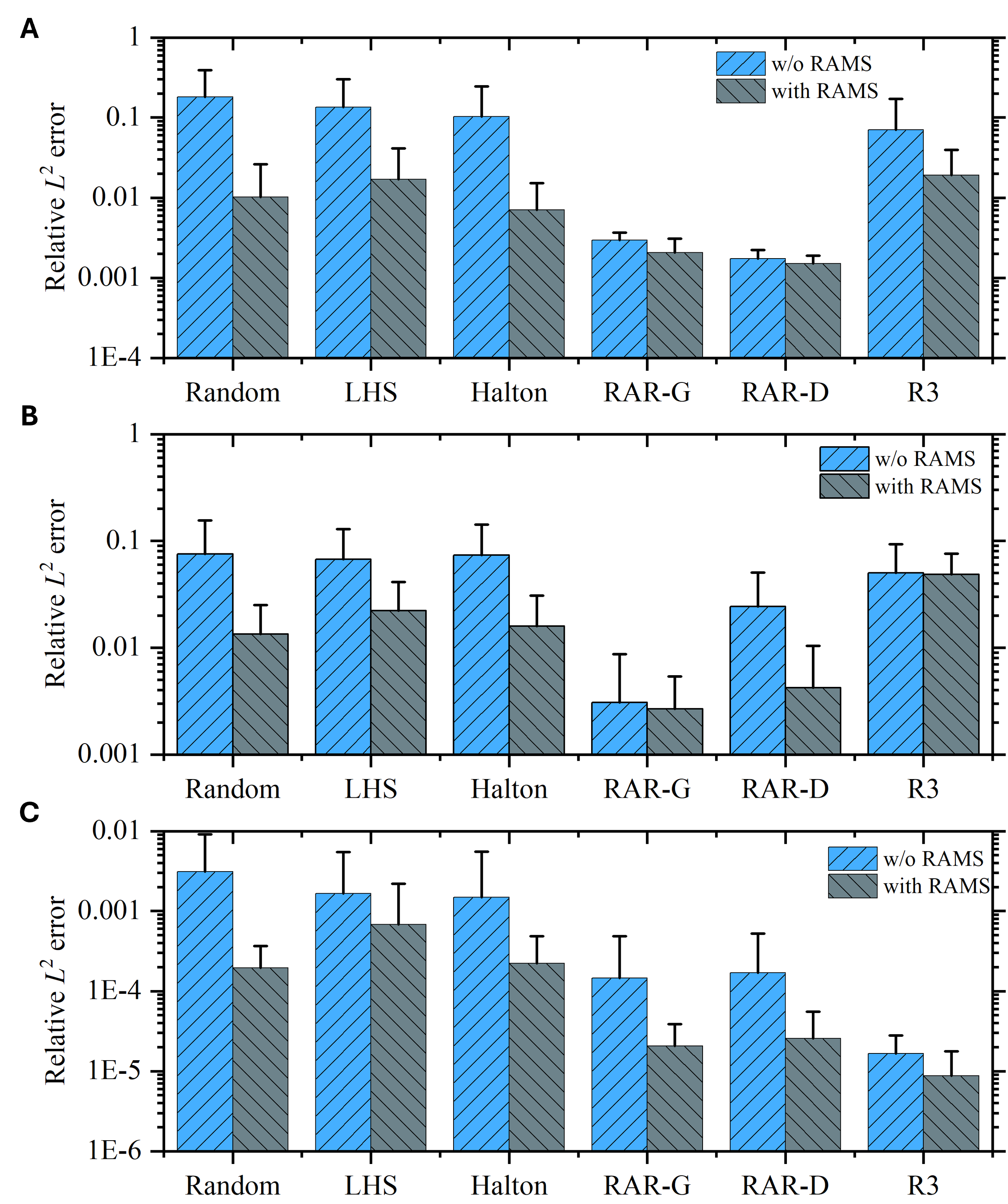}
    \caption{\textbf{Comparison of different sampling methods with and without RAMS for PINN problems.} 
    (\textbf{A}) Burgers' equation.
    (\textbf{B}) Wave equation.
    (\textbf{C}) Poisson equation.
    Vertical error bars denote one standard deviation.}
    \label{fig:ex11_13}
\end{figure}

We visualize the ground-truth solution, the solution from RAR-G with RAMS, and the corresponding pointwise error in Fig.~\ref{fig:pinn_vis}A.
Model performances from different sampling methods are reported via the relative $L^2$ error over 10 independent runs (Fig.~\ref{fig:ex11_13}A). 
RAMS significantly improves the non-adaptive sampling methods, reducing mean errors by more than an order of magnitude. For example, random sampling drops from $0.181$ to $0.010$. RAMS also clearly enhances adaptive sampling methods. For example, for R3, the error decreases by approximately $80\%$.

\subsubsection{1D wave equation}
\label{sec:pinn_wave}

We consider the wave equation:
\begin{equation*}
    \begin{aligned}
        \frac{\partial^2 u}{\partial t^2} = 4\frac{\partial^2 u}{\partial x^2}, & \quad \text{for } x \in [0,1], t \in [0,1], \\
        u(0,t)=u(1,t)=0, & \quad \text{for } t \in [0,1], \\
        \frac{\partial u}{\partial t}(x,0) = 0 ,& \quad \text{for } x \in [0,1], \\
        u(x,0) = \sin(\pi x) + \frac{1}{2}\sin(4\pi x), & \quad \text{for } x \in [0,1],
    \end{aligned}
\end{equation*}
whose solution (Fig.~\ref{fig:pinn_vis}B) has a closed-form \cite{wu2023comprehensive} as
\begin{equation*}
    u(x,t) = \sin(\pi x) \cos(2\pi t) + \frac{1}{2}\sin(4\pi x) \cos(8\pi t).
\end{equation*}

The performances of different sampling methods are illustrated via the relative $L^2$ error over 10 independent runs in Fig.~\ref{fig:ex11_13}B. For all sampling methods, RAMS consistently improves their accuracy, and on average, the improvement exceeds $50\%$. In Fig.~\ref{fig:pinn_vis}B, we show the ground-truth solution, the prediction of MLP from RAR-G with RAMS, and the corresponding pointwise error.

\subsubsection{2D Poisson equation}
\label{sec:pinn_poisson}

We consider a benchmark problem from adaptive finite element tests \cite{mitchell2013collection}: a Poisson equation defined as
\begin{equation*}
    \begin{aligned}
        -\Delta u(x_1,x_2) = s(x_1,x_2), & \quad \text{for } (x_1,x_2) \in \Omega, \\
        u(x_1,x_2) = g(x_1,x_2), & \quad \text{for } (x_1,x_2) \in \partial \Omega,
    \end{aligned}
\end{equation*}
whose domain is $\Omega = [-1,1]^2$. The PDE solution $u$ is selected as
\begin{equation*}
    u(x_1,x_2) = \exp{(-1000[(x_1-0.5)^2 + (x_2-0.5)^2])},
\end{equation*}
which exhibits a sharp interior peak at $(0.5,0.5)$ and decays rapidly away from the peak (Fig.~\ref{fig:pinn_vis}C).

The relative $L^2$ errors over 10 independent runs for different sampling methods are shown in Fig.~\ref{fig:ex11_13}C. RAMS reduces error by approximately 40--95\% across all non-adaptive and adaptive sampling methods, indicating its robust compatibility with different sampling methods. The ground-truth solution and MLP solution from RAR-G with RAMS are plotted with the pointwise error in Fig.~\ref{fig:pinn_vis}C.

\subsubsection{High-dimensional problem}
\label{sec:pinn_hd}

We consider a high-dimensional PDE problem in Ref.~\cite{tang2023das} to evaluate the proposed RAMS method:
\begin{equation*}
    \begin{aligned}
        -\Delta u(\bm{x}) = s(\bm{x}), & \quad \text{for } \bm{x} \in \Omega, \\
        u(\bm{x}) = g(\bm{x}), & \quad \text{for } \bm{x} \in \partial \Omega,
    \end{aligned}
\end{equation*}
where $\Omega=[-1,1]^d$ is the computational domain in a $d$-dimensional space. The exact solution is
\begin{equation*}
    u(\bm{x}) = \exp{(-10\|\bm{x}\|_2^2)}.
\end{equation*}

We apply random sampling with and without RAMS using a total of $|\mathcal{T}|=20,000$ collocation points. For the random sampling with RAMS, $|\mathcal{T}_2|=10,000$ points are trainable, and at each resampling stage, $|\hat{\mathcal{T}}|=2,000$ trainable points are updated by Adam for $n_{RAMS}=100$ epochs. The number of resampling stages and the Adam epochs for model training in each stage are set to $t_r=160$ and $n_{train}=500$, respectively. 
After the final resampling stage, the model is further trained for 20,000 Adam epochs, followed by 300 L-BFGS iterations.

\begin{figure}
    \centering
    \includegraphics[width=\linewidth]{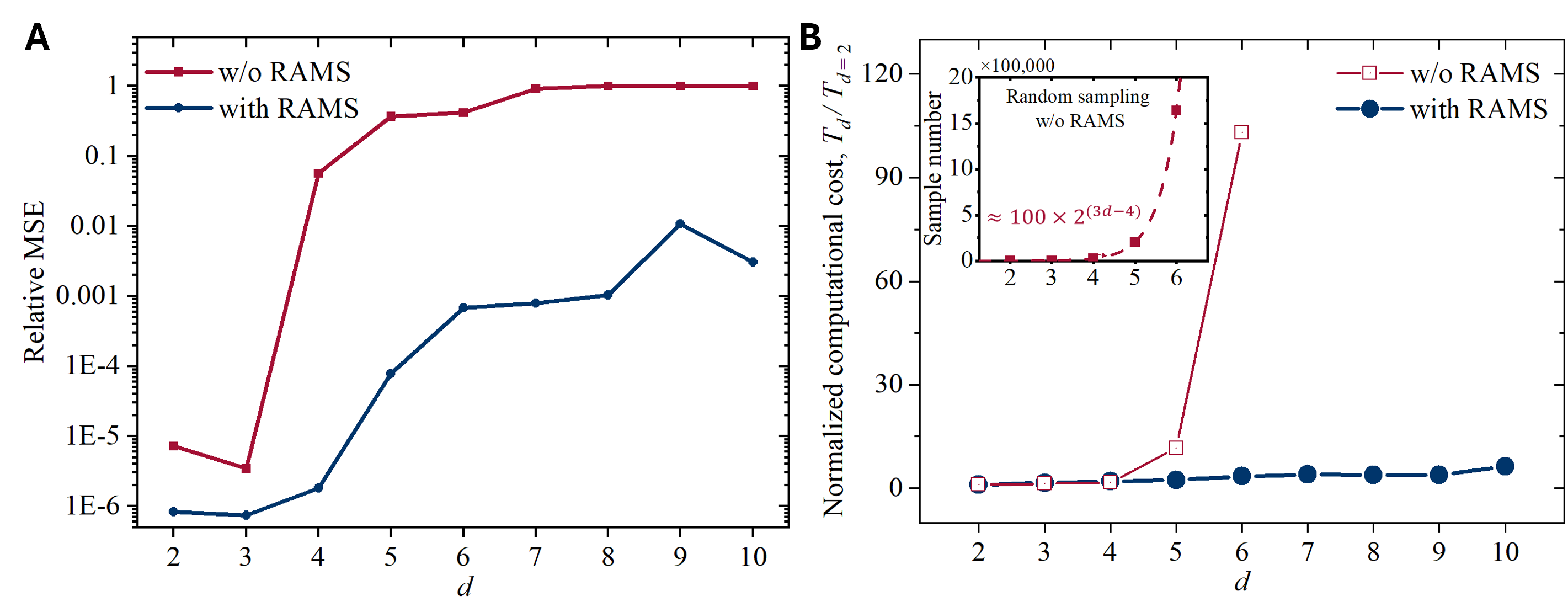}
    \caption{\textbf{RAMS for high-dimensional PDEs.} 
    (\textbf{A}) Relative MSE of random sampling with and without RAMS for different dimensions $d$. 
    (\textbf{B}) Minimal training computational cost $T_d$ required to reach RMSE $\le 10^{-3}$ for each method. The cost is normalized by each method’s cost at $d=2$ ($T_{d=2}$). The inset panel reports the minimal number of collocation points needed by random sampling without RAMS, which grows exponentially with $d$.}
    \label{fig:ex14}
\end{figure}

To assess the method performance, we use the relative mean squared error (RMSE) $E$ computed over two disjoint test sets~\cite{tang2023das}:
\begin{equation*}
E=\frac{\sum_{\xi_i\in\bm{\xi}_1}\!\big(u_{\text{pred}}(\xi_i)-u(\xi_i)\big)^2
+\sum_{\xi_i\in\bm{\xi}_2}\!\big(u_{\text{pred}}(\xi_i)-u(\xi_i)\big)^2}
{\sum_{\xi_i\in\bm{\xi}_1}\!u(\xi_i)^2+\sum_{\xi_i\in\bm{\xi}_2}\!u(\xi_i)^2},
\end{equation*}
where $\bm{\xi}_1$ is a set of 1000 points sampled in hyperspherical coordinates centered at the origin with radial coordinate $r\sim U(0,1)$, and $\bm{\xi}_2$ is a set of 1000 points sampled uniformly from $\Omega$. We show the RMSE for different dimensions $d$ in Fig.~\ref{fig:ex14}A. Without RAMS, random sampling fails to recover the solution as $d$ increases; for $d \geq 7$, the RMSE saturates near 1 ($E\approx1$), indicating failure on these high-dimensional PDEs. In contrast, RAMS enables the random sampling to remain a small error below $10^{-2}$ using the same amount of collocation points even when $d = 10$.

We next estimate the minimal computational cost for training needed to achieve $E\le 10^{-3}$ for two sampling strategies. The training and sampling setups vary with the dimension $d$ as follows.
\begin{itemize}
    \item Collocation points for random sampling without RAMS:
    For each $d$, we train the PINN for 100,000 Adam epochs, followed by 300 L-BFGS iterations. The number of collocation points begins at $|\mathcal{T}|=100$; if the target accuracy is not met, $|\mathcal{T}|$ is doubled and PINN is retrained. This procedure is repeated until $E \le 10^{-3}$.

    \item Training iterations for random sampling with RAMS:
    We use $|\mathcal{T}|=20,000$ collocation points, with $|\mathcal{T}_2|=10,000$ designated as trainable for all $d$. At each resampling stage (every $n_{train}=500$ epochs), a subset of size $|\hat{\mathcal{T}}|=2,000$ is updated. 
    After the final resampling stage, the model is further trained for 10,000 Adam epochs, followed by 300 L-BFGS iterations. For each $d$, we tune the number of resampling stages $t_r$ and the sample-training iterations $n_{RAMS}$ to achieve $E \leq 10^{-3}$. Specifically, we initialize $t_r=10$ and $n_{RAMS}=0$, and if $E \geq 10^{-3}$, we increase $t_r$ by 10 and $n_{RAMS}$ by 20, and repeat until the criterion is met.

\end{itemize}

For each sampling method, computational costs are normalized by the corresponding cost at $d=2$ (denoted by $T_{d=2}$). All costs were measured as the run time on the same machine with a single NVIDIA A40 GPU. The normalized minimum costs for different sampling methods across varying $d$ are shown in Fig.~\ref{fig:ex14}B. We also show that the minimum number of collocation points required for random sampling grows exponentially with $d$. This growth, thereby, drives a rapid increase in computational cost. As for $d=6$, the cost already exceeds $100 \, T_{d=2}$, we stop testing the random sampling for $d\ge 7$ due to prohibitive runtime. By contrast, the cost of random sampling with RAMS scales approximately linearly in $d$: the computational costs at $d=6$ and $d=10$ are about $3.5\,T_{d=2}$ and $6\,T_{d=2}$, respectively, highlighting the efficiency of the proposed RAMS method.

\subsection{Physics-informed operator learning}
\label{sec:ex_piol}

We evaluate RAMS on PI operator learning using two sampling methods: random sampling and RAR-G. During each resampling stage, we update only function samples, and the collocation points for the trunk net input remain fixed. To demonstrate the effectiveness of RAMS, we first test three PDEs: the diffusion-reaction equation (Section~\ref{sec:piol_diff}), the advection equation (Section~\ref{sec:piol_adv}), and the Poisson equation with piecewise-constant conductivity (Section~\ref{sec:piol_poi}). We then test a one-dimensional dynamic system (Section~\ref{sec:piol_dynamic}) on which conventional sampling methods perform poorly \cite{wang2024das2}.

\subsubsection{Diffusion-reaction equation}
\label{sec:piol_diff}

We consider a diffusion-reaction equation with zero initial and boundary conditions:
\begin{equation*}
    \frac{\partial u}{\partial t} = D \frac{\partial^2 u}{\partial t^2} + ku^2+v(x), \quad \text{for } (x,t) \in [0,1] \times [0,1].
\end{equation*}
where $k=10^{-2}$. We use a PI-DeepONet to learn the operator mapping from $v$ to $u$:
\begin{equation*}
    \GG: v \mapsto u,
\end{equation*}
where the input functions $v$ are sampled from a zero-mean GRF with a Gaussian kernel as $v \sim \mathcal{GP}(0, \exp{(\frac{\|x-y\|^2}{2l_{train}^2})})$, where the correlation length of training data $l_{train}$ is sampled randomly from $U(0.1,0.8)$.

\begin{figure}[htbp]
    \centering
    \includegraphics[width=0.7\linewidth]{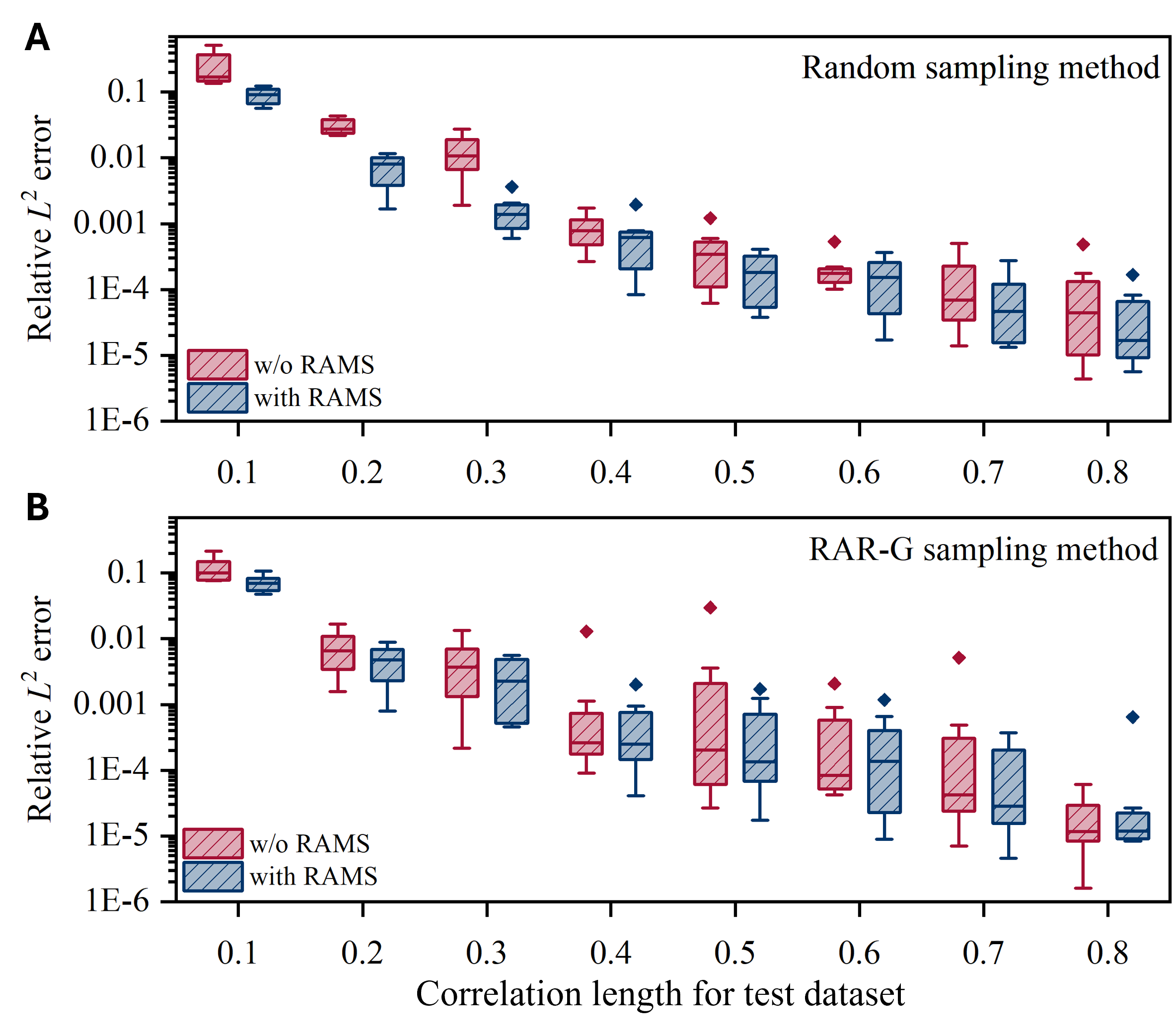}
    \caption{\textbf{PI operator learning of the diffusion-reaction equation.} 
    The correlation length $l_{train}$ of training data is randomly sampled from $U(0.1,0.8)$. Relative $L^2$ errors for different test data correlation lengths $l_{test}$ are computed from eight independent runs.
    (\textbf{A}) Random sampling with and without RAMS.
    (\textbf{B}) RAR-G with and without RAMS.}
    \label{fig:ex21}
\end{figure}

We use the random sampling and RAR-G with and without RAMS for this example. For each method, we evaluate the trained network on eight test datasets with correlation lengths $l_{test} \in \{0.1, 0.2, \dots, 0.8\}$. Performances of different sampling methods are reported via the relative $L^2$ errors on all eight test datasets (Fig.~\ref{fig:ex21}). RAMS yields clear accuracy improvement across different test datasets, especially for those of small correlation lengths. For example, for random sampling, when $l=0.1$, the mean error decreases from approximately $0.25$ (without RAMS) to $0.08$ (with RAMS), i.e., a 70\% error reduction. A representative example of the input function, the ground-truth solution, the PI-DeepONet prediction with RAR-G and RAMS, and the piecewise error are shown in Fig.~\ref{fig:piol_vis}A. Additionally, we show the necessity of the kernel smoothing method as the projector in RAMS in Appendix~\ref{sec:projector_ablation}.

\subsubsection{Advection equation}
\label{sec:piol_adv}

This example aims to investigate the extrapolation capability and computational cost of the proposed RAMS method. We utilize the PI operator learning on the following advection equation tested in Ref.~\cite{wang2021learning}:
\begin{equation*}
    \begin{aligned}
        \frac{\partial s}{\partial t} + u \frac{\partial s}{\partial x} = 0, &\quad \text{for } (x,t) \in (0,1) \times (0,1), \\
    s(x,0) = \sin(\pi x), &\quad \text{for } x \in (0,1), \\
    s(0,t) = \sin(\frac{\pi}{2} t), &\quad \text{for } t \in (0,1). \\
    \end{aligned}
\end{equation*}
A DeepONet is used to approximate the operator mapping from $u$ to $s$ as
\begin{equation*}
    \GG: u \mapsto s.
\end{equation*}
For network training, the input function is sampled from a zero-mean GRF with a Gaussian kernel of a fixed correlation length $l_{train}=0.2$.  

\begin{figure}[htbp]
    \centering
    \includegraphics[width=\linewidth]{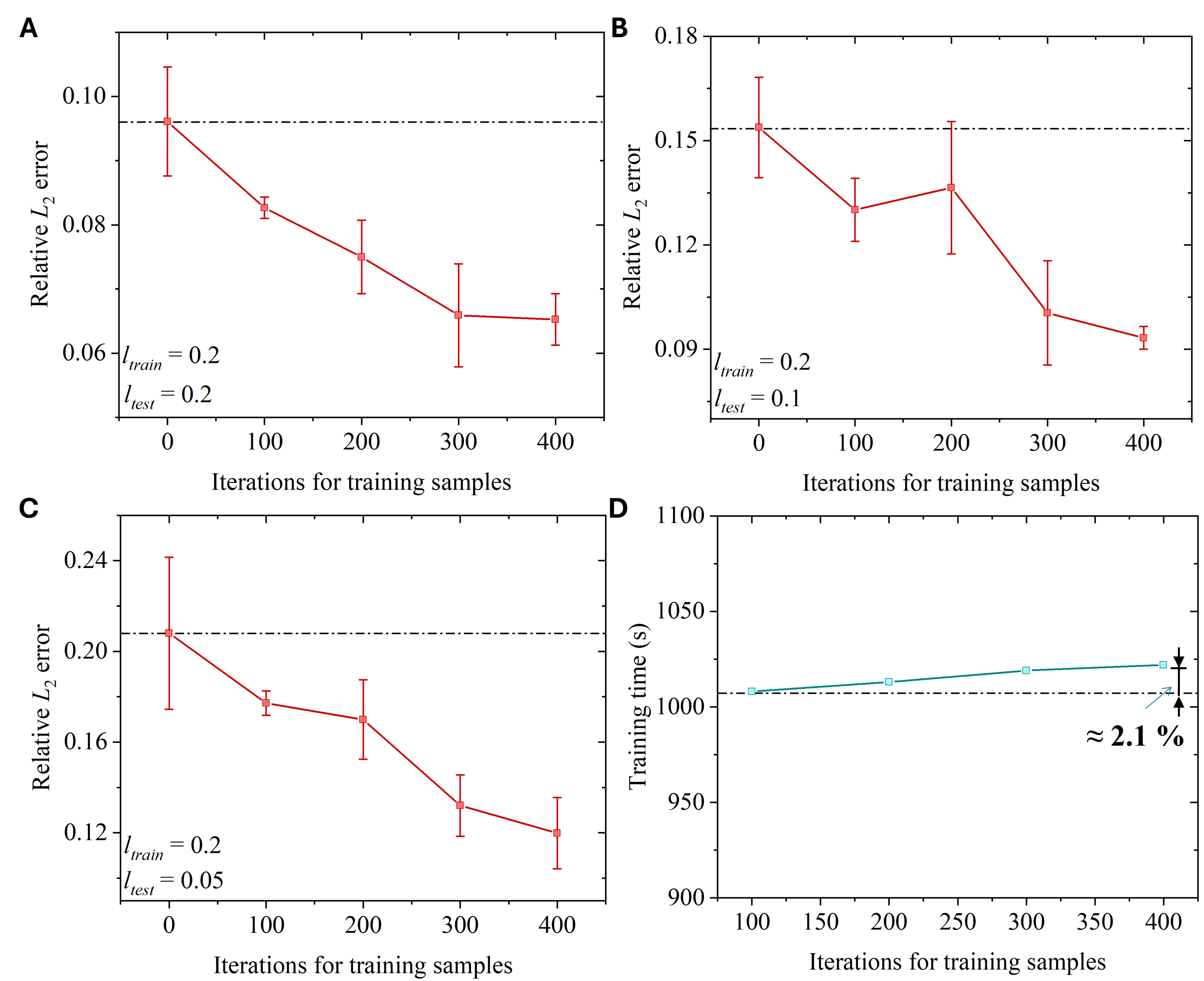}
    \caption{\textbf{PI operator learning of the advection equation.} 
    (\textbf{A}) Relative $L^2$ error on the test dataset with the correlation length $l_{test}=0.2$.
    (\textbf{B}) $l_{test}=0.1$.
    (\textbf{C}) $l_{test}=0.05$. The red lines are the mean values, and vertical bars indicate one standard deviation from eight independent runs. The black dashed lines denote the baseline results of random sampling without RAMS. 
    (\textbf{D}) The training cost versus the sample training iterations in RAMS.
    }
    \label{fig:ex22}
\end{figure}

We apply random sampling with and without RAMS and find that RAMS improves the network generalizability. Specifically, we evaluate the trained PI-DeepONet on three test datasets with input functions sampled from GRFs with correlation lengths $l_{test} \in \{0.2, 0.1, 0.05\}$. The number of RAMS sample-training iterations, $n_{RAMS}$, is varied from 0 to 400. RAMS substantially improves interpolation accuracy: when $l_{test}=l_{train}=0.2$, the error is reduced by approximately $30\%$ (Fig.~\ref{fig:ex22}A). RAMS also enhances extrapolation, yielding roughly $40\%$ lower error on test datasets with $l_{test}=0.1$ (Fig.~\ref{fig:ex22}B) and $0.05$ (Fig.~\ref{fig:ex22}C), demonstrating improvements in both interpolation and extrapolation. In addition, an example of the input function, the ground-truth solution, PI-DeepONet prediction, and its pointwise error are shown in Fig.~\ref{fig:piol_vis}B.

We further observe that the effectiveness of RAMS generally increases with the number of sample training iterations $n_{RAMS}$: larger $n_{RAMS}$ can help in locating regions of higher PDE residuals, leading to a more informative resampling. RAMS introduces additional computational cost, which increases with the sample training iteration $n_{RAMS}$. However, this additional cost is very small compared with network training (Fig.~\ref{fig:ex22}D). Even at $n_{RAMS}=400$, the computational overhead of RAMS is about $2.1\%$. This is because the number of sample parameters is negligible relative to the network parameters.

\subsubsection{Poisson equation}
\label{sec:piol_poi}

We consider a Poisson problem with piecewise-constant conductivity $k$ and zero Dirichlet boundary conditions:
\begin{equation*}
    \begin{aligned}
        \nabla \cdot \left( k \nabla u \right) &= f, \quad \text{for } x \in \Omega, \\
        k &= 
        \begin{cases}
            0.5, \quad \text{for } x \in \Omega', \\
            1.0, \quad \text{for } x \in \Omega \setminus \Omega',
        \end{cases}
    \end{aligned}
\end{equation*}
where $\Omega=[-1,1]^2$ and $\Omega'=[-0.3,0.3]^2$ are two rectangular areas, and $f=\frac{v}{\|v\|}$ is a normalized function sampled via a GRF, $v \sim \mathcal{GP}(0, k_l)$, of the correlation length $l=0.3$. 
A DeepONet is trained to learn the operator mapping from $f$ to $u$:
\[
\mathcal{G}: f \mapsto u.
\]

\begin{figure}[htbp]
    \centering
    \includegraphics[width=\linewidth]{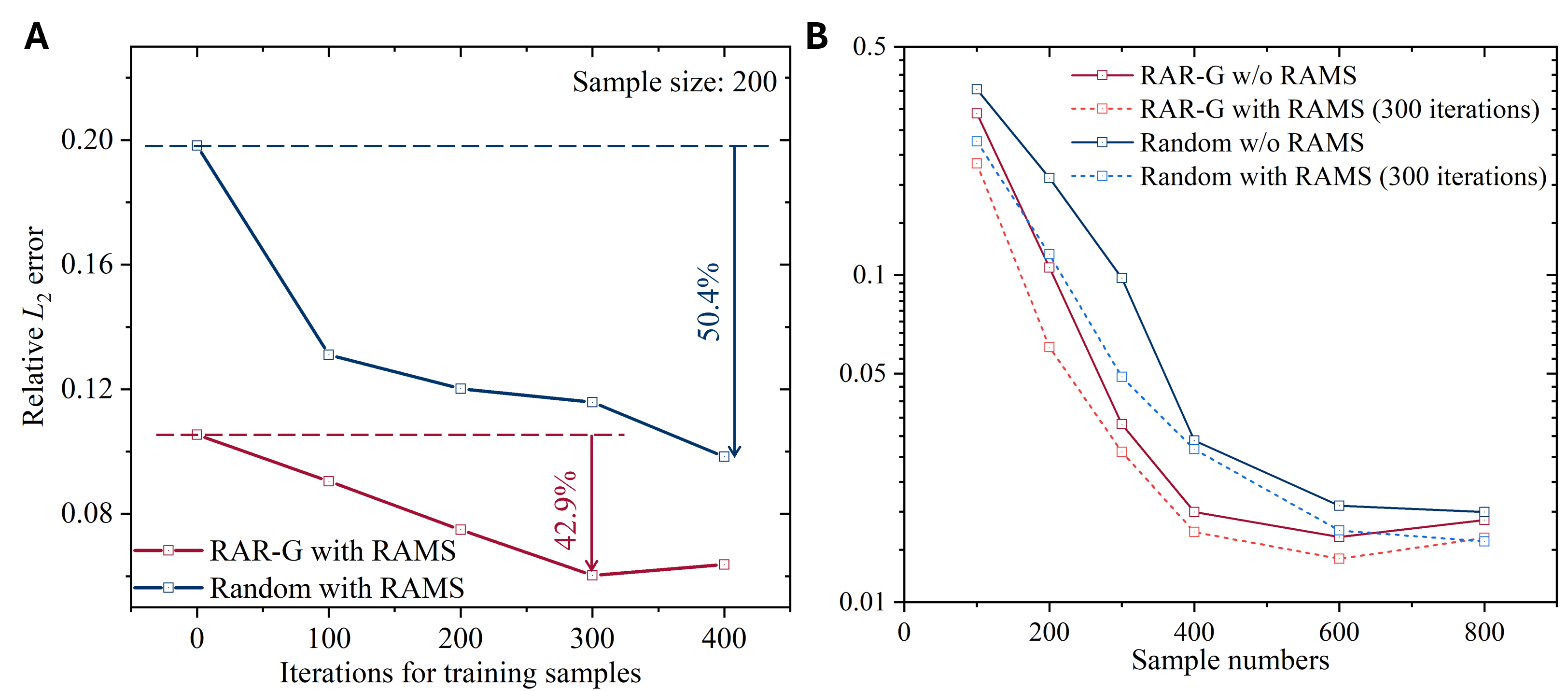}
    \caption{\textbf{PI operator learning of the Poisson equation.} (\textbf{A}) Relative $L^2$ error for different sampling methods with different sample-training iterations in RAMS.
    (\textbf{B}) Relative $L^2$ error for different sampling methods with different training dataset sizes.}
    \label{fig:ex23}
\end{figure}

We evaluate random sampling and RAR-G, each with and without RAMS. For random sampling, we use $n_{sam}$ samples, while for RAR-G, we gradually increase the sample size to $n_{sam}$. Fig.~\ref{fig:ex23}A shows the performances of different sampling methods versus the sample-training iterations in RAMS, $n_{RAMS}$, when fixing $n_{sam}=200$. RAMS reduces error by more than 40\% for both random sampling and RAR-G, demonstrating its robustness across sampling strategies. However, performance does not always improve with additional iterations: for RAR-G with RAMS, the best result occurs at $n_{RAMS}=300$ rather than $n_{RAMS}=400$, underscoring the importance of tuning $n_{RAMS}$.

Next, we vary $n_{sam}$ from 100 to 800 while fixing the sample-training iterations in RAMS at $n_{RAMS}=300$. RAMS leads to substantial improvements across all sample sizes (Fig.~\ref{fig:ex23}B), over $30\%$ at $n_{sam}=100$ and about $20\%$ at $n_{sam}=800$, indicating stable performances of the RAMS method irrespective of training dataset size. In addition, we visualize a representative ground-truth solution, the corresponding PI-DeepONet prediction from RAR-G with RAMS, and the associated piecewise error in Fig.~\ref{fig:piol_vis}C.

\subsubsection{One-dimensional dynamic system}
\label{sec:piol_dynamic}

We consider a dynamic system where both random sampling and the vanilla RAR-G fail to generate accurate solutions efficiently \cite{wang2024das2}:
\begin{equation*}
    \frac{du}{dx} = \exp\left(-D \|\bm{\xi} - \bm{0.5}\|^2\right) f(x, \bm{\xi}), \quad \text{for } x \in [0,1],
\end{equation*}
where $D=6$ is a constant parameter, and $\bm{\xi} \in [-1,1]^d$ is a coefficient vector for the source term with dimension $d = 8$. The boundary condition is given by
\begin{equation*}
    u(x) = 0, \quad \text{for } x=0.
\end{equation*}
The function $f$ is parameterized by a space spanned by orthogonal Chebyshev polynomials of the first kind. Denoting these polynomials by $f_i$, where $i$ denotes the degree, the polynomial space of degree $d$ is defined as
\begin{equation*}
    \mathcal{V} = \left\{ f(x, \bm{\xi}) = \sum_{i=0}^{d-1} \xi_i f_i(x) : |\xi_i| \leq 1 \right\}.
\end{equation*}

\begin{figure}[htbp]
    \centering
    \includegraphics[width=0.8\linewidth]{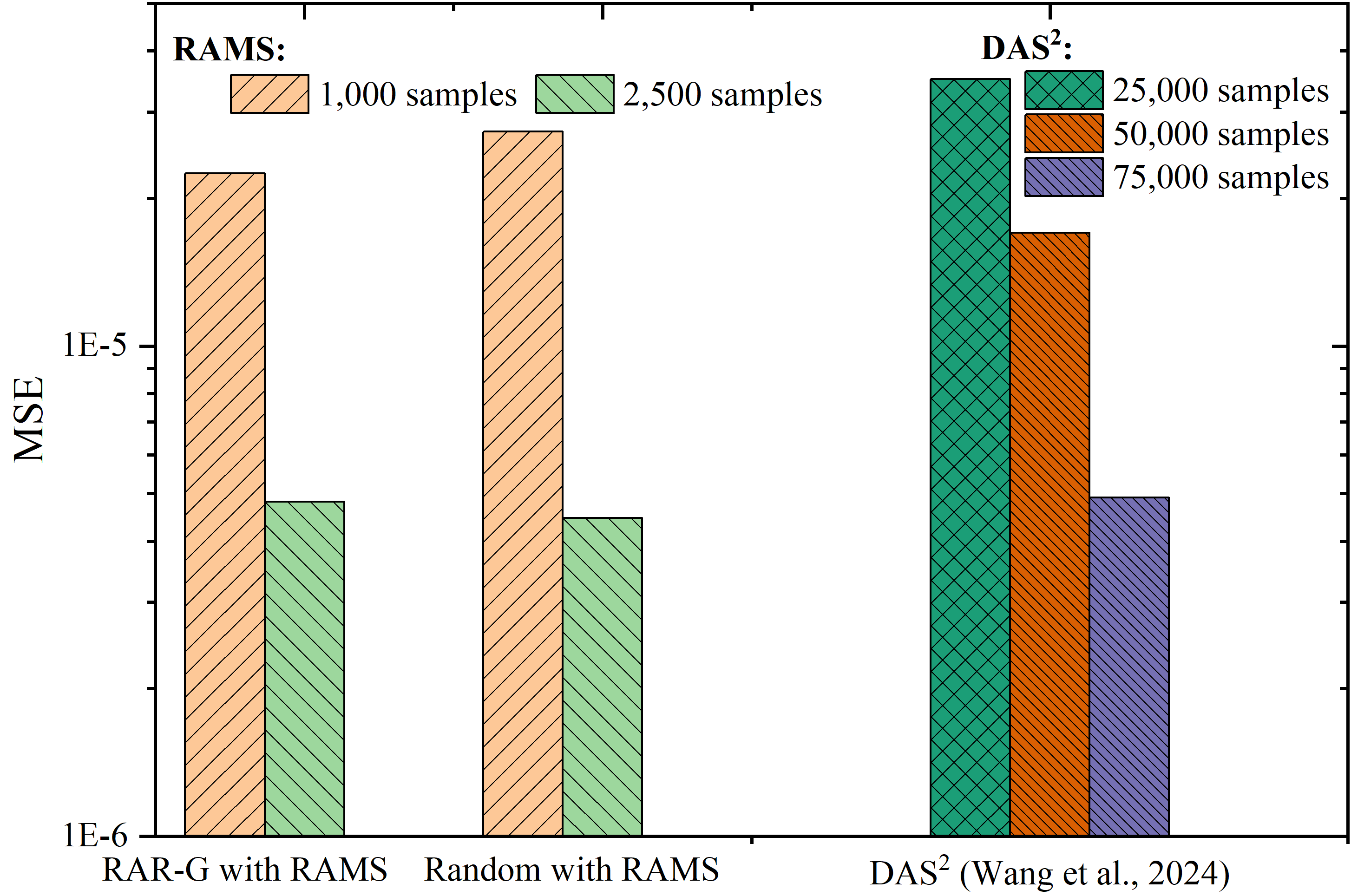}
    \caption{\textbf{PI operator learning of a dynamic system.} 
    RAR-G with RAMS and random sampling with RAMS achieve comparable accuracy to DAS$^2$~\cite{wang2024das2} while using significantly fewer samples. 1,000 and 2,500 samples in our methods match the performance of DAS$^2$ trained with 25,000 and 75,000 samples, respectively.
    }
    \label{fig:ex24}
\end{figure}

We test random sampling and RAR-G with RAMS. To align with the experimental setup of Ref.~\cite{wang2024das2}, we use the same model architecture, training procedure, and sampling setup, differing only in the inclusion of RAMS.
To evaluate the performance of the trained DeepONet, we use a test dataset consisting of 10,000 $\bm{\xi}$ samples, uniformly sampled from a $d$-dimensional ball centered at $\bm{0.5}$ with a radius of 0.5. We compare RAR-G with RAMS and random sampling with RAMS with DAS\textsuperscript{2}~\cite{wang2024das2} in Fig.~\ref{fig:ex24}. While DAS\textsuperscript{2} can achieve accurate solutions, it requires a prohibitively large number of samples. In contrast, our RAMS method achieves similar accuracy with only about 3\% of the samples required by DAS\textsuperscript{2}. Specifically, 1,000 and 2,500 samples in our method match the performance of DAS\textsuperscript{2} trained with 25,000 and 75,000 samples, respectively. This significant reduction in sample size underscores the efficiency and effectiveness of RAMS in operator learning.

\subsection{Data-driven operator learning}
\label{sec:ex_ddol}

The effectiveness of the proposed RAMS method in PI machine learning, including both PINN and PI operator learning, has been demonstrated in previous examples. Here, we further illustrate the performance of our RAMS method in data-driven operator learning applied to wave equation with discontinuous velocity (Section~\ref{sec:ddol_wave}) and 2D Burgers' equation (Section~\ref{sec:ddol_burgers}).

\subsubsection{Wave equation with discontinuous velocity}
\label{sec:ddol_wave}

We consider the wave equation with discontinuous velocity:
\begin{equation*}
    \begin{aligned}
        \frac{\partial^2 u}{\partial t^2} - c^2(x)\frac{\partial^2 u}{\partial x^2} = 0, & \quad \text{for } x \in [0,1], t \in [0,4], \\
        u(0,t)=u(1,t)=0, & \quad \text{for } t \in [0,4], \\
        \frac{\partial u}{\partial t}(x,0) = 0 ,& \quad \text{for } x \in [0,1], \\
        u(x,0) = v(x), & \quad \text{for } x \in [0,1],
    \end{aligned}
\end{equation*}
where the velocity $c$ is discontinuous at $x=0.7$:
\begin{equation*}
    c(x) = 
        \begin{cases}
            1.0, \quad \text{for } 0 \leq x < 0.7, \\
            0.5, \quad \text{for } 0.7 \leq x \leq 1.0.
        \end{cases}
\end{equation*}
We employ a DeepONet to learn the operator mapping from the initial condition to the solution:
\begin{equation*}
    \mathcal{G}: v \mapsto u,
\end{equation*}
where $v$ is generated from $v'$ via $v(x) = x(1-x)v'(x)$, and $v'$ is sampled from a GRF with an Gaussian kernel of correlation length $l=0.3$.

The vanilla random sampling and RAR-G with RAMS are utilized to generate the training datasets. The total training sample sizes, $n_{sam}$, are varied from 50 to 200. In RAR-G with RAMS, the iterations for sample-training, $n_{RAMS}$, in each resampling stage are varied from 0 to 400. Detailed setups for different sampling methods can be found in Appendix~\ref{apd:setup}.

\begin{figure}[htbp]
    \centering
    \includegraphics[width=\linewidth]{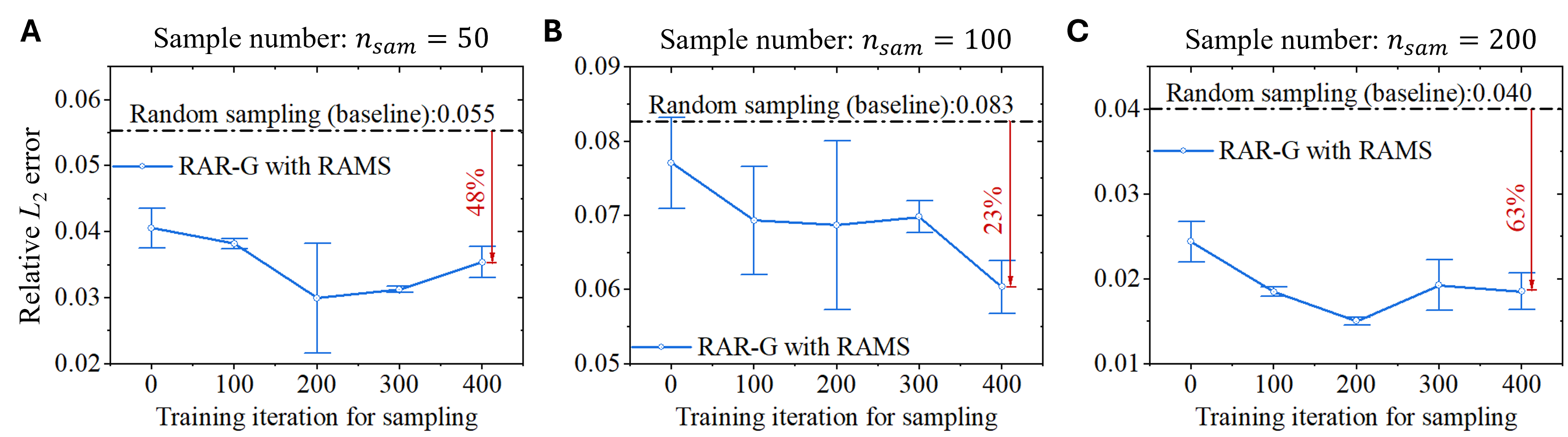}
    \caption{\textbf{Data-driven operator learning of the wave equation with discontinuous velocity.} 
    (\textbf{A}) Relative $L^2$ error for DeepONets trained with 50 samples.
    (\textbf{B}) Relative $L^2$ error for DeepONets trained with 100 samples.
    (\textbf{C}) Relative $L^2$ error for DeepONets trained with 200 samples.
    The blue lines are the mean errors, and vertical bars indicate one standard deviation from three independent runs of RAR-G with RAMS.
    The black dashed lines correspond to the mean errors from the random sampling method.}
    \label{fig:ex31}
\end{figure}

The mean relative $L^2$ errors for different settings are shown in Fig.~\ref{fig:ex31}.
The RAMS method achieves significant accuracy improvements, up to 63\%, using the same number of training samples as the baseline method (random sampling). 
This substantial improvement confirms the effectiveness of RAMS in data-driven operator learning.
We also observe that the error reduction from RAMS does not necessarily increase with the number of sampling–training iterations $n_{RAMS}$. 
For example, when $n_{sam}=100$ (Fig.~\ref{fig:ex31}B), the best performance occurs at $n_{RAMS}=200$ rather than $n_{RAMS}=400$, underscoring the need to tune $n_{RAMS}$. 
Finally, the effectiveness of RAMS increases with the total training-sample size: the error reduction rises from $23\%$ at $n_{sam}=50$ (Fig.~\ref{fig:ex31}A) to $63\%$ at $n_{sam}=200$ (Fig.~\ref{fig:ex31}C), suggesting that more samples allow the physics residual to more accurately reflect the model’s training adequacy.
In addition, a representative example of an input function, ground-truth solution, DeepONet prediction from RAR-G with RAMS, and pointwise error are shown in Fig.~\ref{fig:ddol_wave}.

\subsubsection{Two-dimensional Burgers' equation}
\label{sec:ddol_burgers}

We consider the two-dimensional Burgers' equation \cite{kundu2020global} given by
\begin{equation*}
    \begin{aligned}
        \frac{\partial u}{\partial t} + u (\frac{\partial u}{\partial x_1} + \frac{\partial u}{\partial x_2}) = \nu \Delta u, &\quad \text{for } (x_1,x_2) \in \Omega, t \in [0,1], \\
        u(x_1,x_2,0) = v(x_1,x_2), &\quad \text{for } (x_1,x_2) \in \Omega, \\
        u(x_1,x_2,t) = 0, &\quad \text{for } (x_1,x_2) \in \partial \Omega, t \in [0,1],\\
    \end{aligned}
\end{equation*}
where $\nu$ is the viscosity coefficient, and $\Omega = [-1,1]^2$ is the computational domain. A DeepONet is trained to approximate the operator mapping from the initial condition to the solution as
\begin{equation*}
    \GG: v \mapsto u,
\end{equation*}
where $v$ is generated from $v'$ via $v(x_1,x_2) = x_1(1-x_1)x_2(1-x_2)v'(x_1,x_2)$, and $v'$ is sampled from a GRF with a Gaussian kernel of correlation length $l$.

We consider two cases, differentiated by their viscosity coefficients, correlation lengths, and training sample sizes. These variations represent different complexities of the operator learning problem.
\begin{itemize}
    \item Case 1 (small viscosity and sample size): The viscosity coefficient and correlation length are set to $\nu=0.02$ and $l=0.3$, respectively. The sample size $n_{sam}$ ranges from 600 to 1,200.
    \item Case 2 (large viscosity and sample size): The viscosity coefficient and correlation length are set to $\nu=0.1$ and $l=0.5$, respectively. The sample size $n_{sam}$ ranges from 1,500 to 3,000.
\end{itemize}

We compare the vanilla random sampling method and the RAR-G with RAMS across different settings. For RAR-G with RAMS, $n_{ini}=(1-p)n_{sam}$ samples with $p \in \{0.1,0.2,0.4\}$ are initially generated. In each resampling stage of RAMS, $m=\frac{p}{t_r}n_{sam}$ additional samples chosen from $M=1000$ candidates are added to the training dataset, where $t_r \in \{1, 2 \}$ represents the number of resampling stages. In RAMS, the samples are updated using the Adam optimizer for $n_{RAMS}=100$ iterations at each resampling stage.

\begin{figure}[htbp]
    \centering
    \includegraphics[width=0.8\linewidth]{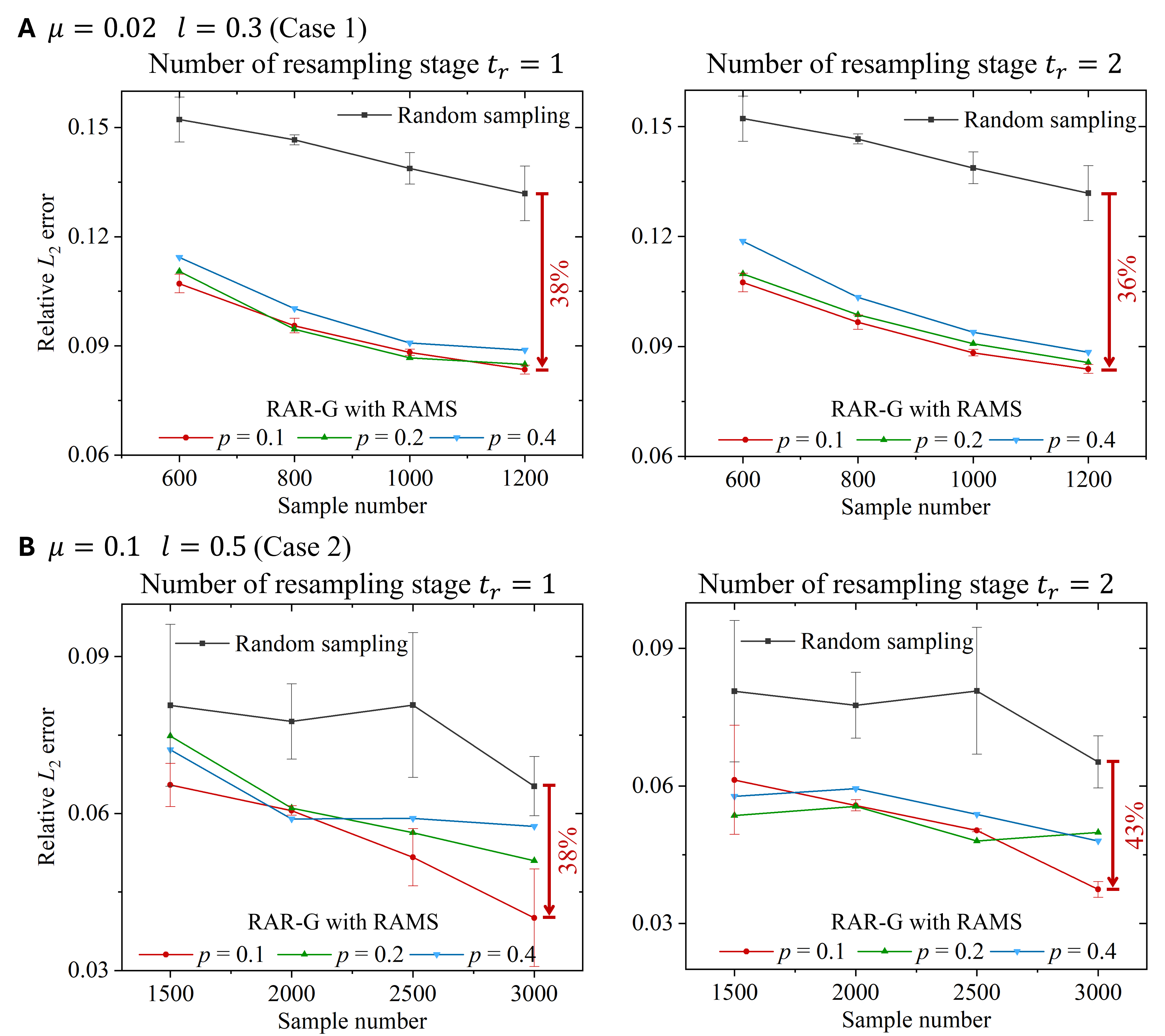}
    \caption{\textbf{Data-driven operator learning of the two-dimensional Burgers' equation.} 
    (\textbf{A}) Relative $L^2$ errors of Case 1 for different trainable sample sizes $p$ and numbers of resampling stages $t_r$.
    (\textbf{B}) Relative $L^2$ errors of Case 2.
    For clarity, the one standard deviation from three independent runs is shown only for the random sampling and RAR-G with RAMS of $p=0.1$.}
    \label{fig:ex32}
\end{figure}

The performance of different sampling methods are evaluated on the mean relative $L_2$ error (Fig.~\ref{fig:ex32}). Across all settings, the largest error reduction occurs at $p=0.1$, suggesting that allocating more samples early helps the model capture the underlying physics and makes the physics residuals more indicative of training adequacy.
The robustness of RAMS is further demonstrated by its consistent performance, reducing errors by approximately 40\% in both a complex case (Case 1; Fig.~\ref{fig:ex32}A) and a simpler one (Case 2; Fig.~\ref{fig:ex32}B).
Additionally, Fig.~\ref{fig:ddol_burgers} shows a representative example of a ground-truth solution, the prediction of a DeepONet from RAR-G with RAMS, and the associated pointwise error.

%% file: content/conclusion.tex
\section{Conclusions}
\label{sec:con}

Machine learning (ML) has become an essential tool for solving PDE problems. The efficiency of ML in solving PDEs often depends on the choice of sampling methods. In this study, we introduce a novel RAMS strategy, where samples are treated as trainable parameters. These samples are optimized to maximize the physics residual using gradient-based optimization methods, effectively identifying areas that require additional training.

The proposed RAMS method demonstrates significant improvements in enhancing various existing sampling strategies across different tasks, including PINN, PI operator learning, and data-driven operator learning. While the current work focuses on solving one PDE system, future studies will explore the application of RAMS in training foundation models \cite{liu2024prose} to solve multiple PDEs.

%% file: content/app_model.tex
\section{Network architectures}
\label{apd:models}

We provide the network architectures used in all the examples. The hyperbolic tangent function is used as the activation function for all the models employed.
Table~\ref{tab:pinn_mlp} summarizes the network architectures for the PINN problems in Section~\ref{sec:ex_pinn}. Table~\ref{tab:deeponets} lists the DeepONet architectures for the operator learning problems in Sections~\ref{sec:ex_piol}--\ref{sec:ex_ddol}. In Section~\ref{sec:piol_adv}, the DeepONet branch and trunk networks adopt the modified MLP architecture proposed in Ref.~\cite{wang2021learning} with better performance.

\begin{table}[htbp]
\centering
\caption{\textbf{Network architectures for PINN problems.} MLP $(a \times b)$ denotes a MLP with $a$ hidden layers and $b$ neurons per hidden layer.}
\label{tab:pinn_mlp}
\begin{tabular}{ll}
  \toprule
  Problem & Network \\ \midrule
  Section~\ref{sec:pinn_burgers} 1D Burgers' equation & MLP ($3 \times 100$) \\
  Section~\ref{sec:pinn_wave} 1D wave equation           & MLP ($3 \times 100$) \\
  Section~\ref{sec:pinn_poisson} 2D Poisson equation     & MLP ($3 \times 100$) \\
  Section~\ref{sec:pinn_hd} High-dimensional PDE      & MLP ($3 \times 100$) \\ \bottomrule
\end{tabular}
\end{table}

\begin{table}[htbp]
\centering
\caption{\textbf{DeepONet architectures for operator learning problems.} MLP $(a \times b)$ denotes a MLP with $a$ hidden layers and $b$ neurons per hidden layer.
CNNs comprise three 32-channel convolutional layers (kernel size 5, stride 2), followed by two 128-neuron fully-connected layers.
}
\label{tab:deeponets}
\begin{tabular}{lll}
  \toprule
  Problem & Branch net & Trunk net \\
  \midrule
  Section~\ref{sec:piol_diff} Diffusion-reaction equation & MLP ($3 \times 150$) & MLP ($3 \times 150$) \\
  Section~\ref{sec:piol_adv} Advection equation           & Modified MLP ($3 \times 150$) & Modified MLP ($3 \times 150$) \\
  Section~\ref{sec:piol_poi} Poisson equation             & CNN & MLP ($3 \times 350$) \\
  Section~\ref{sec:piol_dynamic} Dynamic system           & MLP ($3 \times 50$)  & MLP ($3 \times 50$)  \\
  Section~\ref{sec:ddol_wave} Wave equation             & MLP ($5 \times 150$) & MLP ($5 \times 150$) \\
  Section~\ref{sec:ddol_burgers} 2D Burgers' equation     & CNN & MLP ($4 \times 350$) \\
  \bottomrule
  \end{tabular}
\end{table}

%% file: content/app_setup_training_sampling.tex
\section{Sampling and training hyperparameters}
\label{apd:setup}

This section summarizes the training and sampling configurations used in Section~\ref{sec:ex}. For network training, we employ Adam with a learning rate of $10^{-3}$ and L-BFGS. For sample training in RAMS, we use Adam with a learning rate of $10^{-2}$. Other setups for network training and sampling are summarized below.
\begin{itemize}
    \item \textbf{PINN in Sections~\ref{sec:pinn_burgers}--\ref{sec:pinn_poisson}:}
    The PINNs for the Burgers' equation (Section~\ref{sec:pinn_burgers}), the wave equation (Section~\ref{sec:pinn_wave}), and the Poisson equation (Section~\ref{sec:pinn_poisson}) use the same setups for network training and sampling.
    For all sampling methods, the number of resampling stages and the Adam epochs per stage are set to $t_r = 75$ and $n_{train} = 200$, respectively. After the final resampling stage, the model is further trained for 5,000 Adam epochs, followed by 1,000 L-BFGS iterations. The other hyperparameters for sampling are listed in Table~\ref{tab:pinn_sampling_setup}.
    
    \item \textbf{PI operator learning for the diffusion-reaction equation in Section~\ref{sec:piol_diff}:}
    For all sampling methods, the number of resampling stages and the Adam epochs per stage are set to $t_r = 50$ and $n_{train} = 1,000$, respectively. After the final resampling stage, the model is further trained for 10,000 Adam epochs, followed by 2,000 L-BFGS iterations. The collocation points for each function are in a $100 \times 100$ uniform grid. The other hyperparameters for sampling are listed in Table~\ref{tab:ex21_sampling_setup}.
    
    \item \textbf{PI operator learning for the advection equation in Section~\ref{sec:piol_adv}:}
    For all sampling methods, the number of resampling stages and the Adam epochs per stage are set to $t_r = 20$ and $n_{train} = 5,000$, respectively. After the final resampling stage, the model is further trained for 50,000 Adam epochs, followed by 2,000 L-BFGS iterations. All function samples use the same set of 10,000 collocation points, uniformly sampled over the domain. The other hyperparameters for sampling are listed in Table~\ref{tab:ex22_sampling_setup}.
    
    \item \textbf{PI operator learning for the Poisson equation in Section~\ref{sec:piol_poi}:}
    For all sampling methods, the number of resampling stages and the Adam epochs per stage are set to $t_r = 25$ and $n_{train} = 5,000$, respectively. After the final resampling stage, the model is further trained for 25,000 Adam epochs, followed by 2,000 L-BFGS iterations. All function samples use the same set of 2,500 collocation points, uniformly sampled over the domain. The other hyperparameters for sampling are listed in Table~\ref{tab:ex23_sampling_setup}.

    \item \textbf{PI operator learning for the dynamic system in Section~\ref{sec:piol_dynamic}:}
    For all sampling methods, the number of resampling stages and the Adam epochs per stage are set to $t_r = 30$ and $n_{train} = 500$, respectively. All function samples use the same set of 100 collocation points, uniformly sampled over the domain. The other hyperparameters for sampling are listed in Table~\ref{tab:ex24_sampling_setup}.

    \item \textbf{Data-driven operator learning for the wave equation in Section~\ref{sec:ddol_wave}:}
    In RAR-G with RAMS, the number of resampling stages is fixed at $t_r=2$.
    In RAMS, the number of sample-training iterations $n_{RAMS}$ is varied from 0 to 400. The other hyperparameters for sampling are listed in Table~\ref{tab:ex31_sampling_setup}.
\end{itemize}

\begin{table}[htbp]
\centering
\caption{\textbf{Sampling setups in Sections~\ref{sec:pinn_burgers}--\ref{sec:pinn_poisson}.}}
\label{tab:pinn_sampling_setup}
\begin{tabular}{ll}
  \toprule
  Sampling method & Hyperparameters \\
  \midrule
Random/LHS/Halton & $|\mathcal{T}|=5,000, \,|\mathcal{T}_1|=4,500, \,|\mathcal{T}_2|=500, \,|\hat{\mathcal{T}}|=50, \,n_{RAMS} = 10$ \\
RAR-G/RAR-D       & $n_{ini}=2,500, \, M=1,000, \,m = 40, \,n_{RAMS} = 5$                                     \\
R3                & $|\mathcal{T}|=2,500, \, n_{RAMS} = 5$ \\
  \bottomrule
  \end{tabular}
\end{table}

\begin{table}[htbp]
\centering
\caption{\textbf{Sampling setups in Section~\ref{sec:piol_diff}.}}
\label{tab:ex21_sampling_setup}
\begin{tabular}{ll}
  \toprule
  Sampling method & Hyperparameters \\
  \midrule
Random                                 & $ |\mathcal{T}|=600, \, |\mathcal{T}_1|=300, \,  |\mathcal{T}_2|=300, \,|\hat{\mathcal{T}}|=60, \, n_{RAMS} = 50$ \\
RAR-G                                  & $ n_{ini}=800, \, M=100, \,m = 8, \,n_{RAMS} = 50$   \\
  \bottomrule
  \end{tabular}
\end{table}

\begin{table}[htbp]
\centering
\caption{\textbf{Sampling setups in Section~\ref{sec:piol_adv}.}}
\label{tab:ex22_sampling_setup}
\begin{tabular}{ll}
  \toprule
  Sampling method & Hyperparameters \\
  \midrule
Random                                 & $ |\mathcal{T}|=200, \, |\mathcal{T}_1|=180, \,  |\mathcal{T}_2|=20, \,|\hat{\mathcal{T}}|=2,  \, n_{RAMS} \in \{0, 100, 200, 300, 400\}$ \\
  \bottomrule
  \end{tabular}
\end{table}

\begin{table}[htbp]
\centering
\caption{\textbf{Sampling setups in Section~\ref{sec:piol_poi}.}}
\label{tab:ex23_sampling_setup}
\begin{tabular}{ll}
  \toprule
  Sampling method & Hyperparameters \\
  \midrule
Random                                 & $ |\mathcal{T}|=n_{sam}, \, |\mathcal{T}_1|=0.5n_{sam}, \,  |\mathcal{T}_2|=0.5n_{sam}, \,|\hat{\mathcal{T}}|=0.05n_{sam}$ \\
RAR-G                                  & $ n_{ini}=0.7n_{sam}, \, M=0.3n_{sam}, \,m = 0.02n_{sam}$   \\
  \bottomrule
  \end{tabular}
\end{table}

\begin{table}[htbp]
\centering
\caption{\textbf{Sampling setups in Section~\ref{sec:piol_dynamic}.} Two sample sizes, $n_{sam} \in \{1000\,, 2500\}$, are considered. }
\label{tab:ex24_sampling_setup}
\begin{tabular}{ll}
  \toprule
  Sampling method & Hyperparameters \\
  \midrule
Random                                 & $ |\mathcal{T}|=n_{sam}, \, |\mathcal{T}_1|=|\mathcal{T}_2|=0.5n_{sam}, \,  |\hat{\mathcal{T}}|=0.05n_{sam}, \, n_{RAMS} = 400$ \\
RAR-G                                  & $ n_{ini}=0.4n_{sam}, \, M=1000, \,m = 0.02n_{sam}, \, n_{RAMS} = 400$   \\
  \bottomrule
  \end{tabular}
\end{table}

\begin{table}[htbp]
\centering
\caption{\textbf{Sampling setups in Section~\ref{sec:ddol_wave}.} Three sample sizes, $n_{sam} \in \{50,100,200 \}$, are considered.}
\label{tab:ex31_sampling_setup}
\begin{tabular}{ll}
  \toprule
  Sampling method & Hyperparameters \\
  \midrule
Random without RAMS                              & $ |\mathcal{T}|=n_{sam}$ \\
RAR-G with RAMS                                 & $ n_{ini}=0.6n_{sam}, \, M=1000, \,m = 0.2n_{sam}$   \\
  \bottomrule
  \end{tabular}
\end{table}

%% file: content/app_vis.tex
\section{Visualization of results}
\label{apd:vis}

\begin{figure}[htbp]
    \centering
    \includegraphics[width=.9\linewidth]{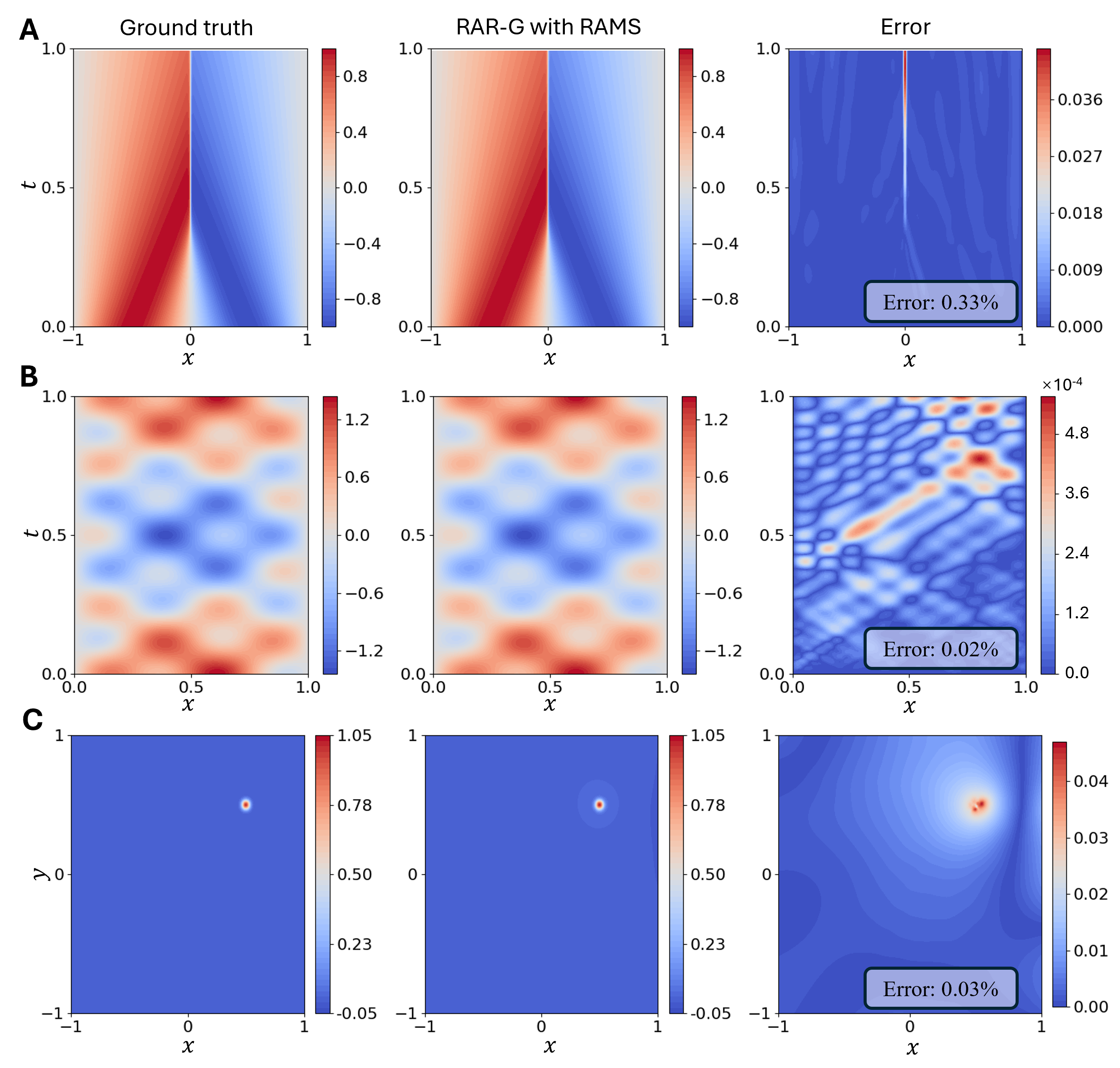}
    \caption{\textbf{Visualization of PINN results.} 
    (\textbf{A}) Burgers' equation.
    (\textbf{B}) Wave equation.
    (\textbf{C}) Poisson equation.}
    \label{fig:pinn_vis}
\end{figure}

\begin{figure}[htbp]
    \centering
    \includegraphics[width=\linewidth]{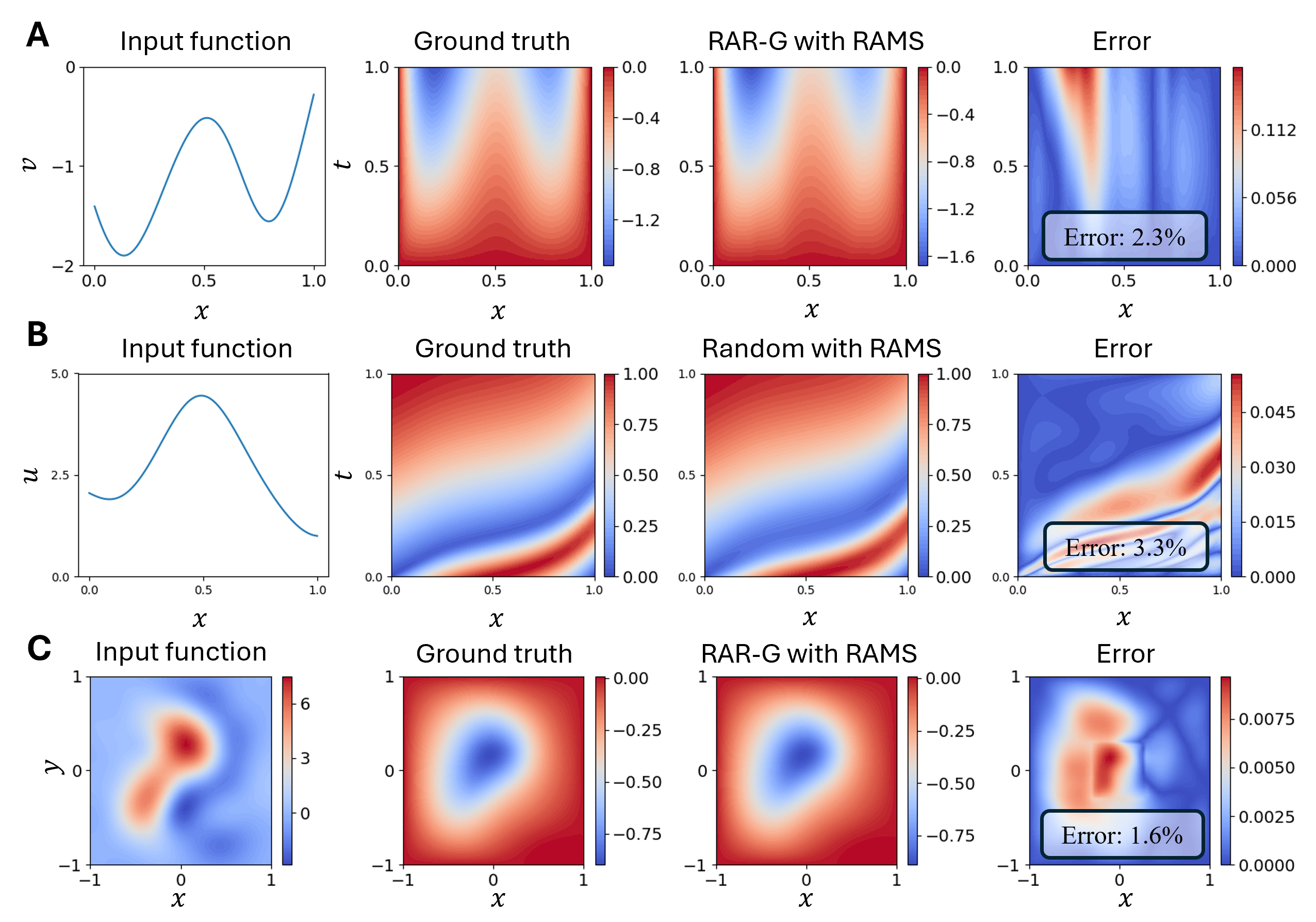}
    \caption{\textbf{Representative examples of PI operator learning.} 
    (\textbf{A}) Diffusion-reaction equation.
    (\textbf{B}) Advection equation.
    (\textbf{C}) Poisson equation.}
    \label{fig:piol_vis}
\end{figure}

\begin{figure}[htbp]
    \centering
    \includegraphics[width=0.6\linewidth]{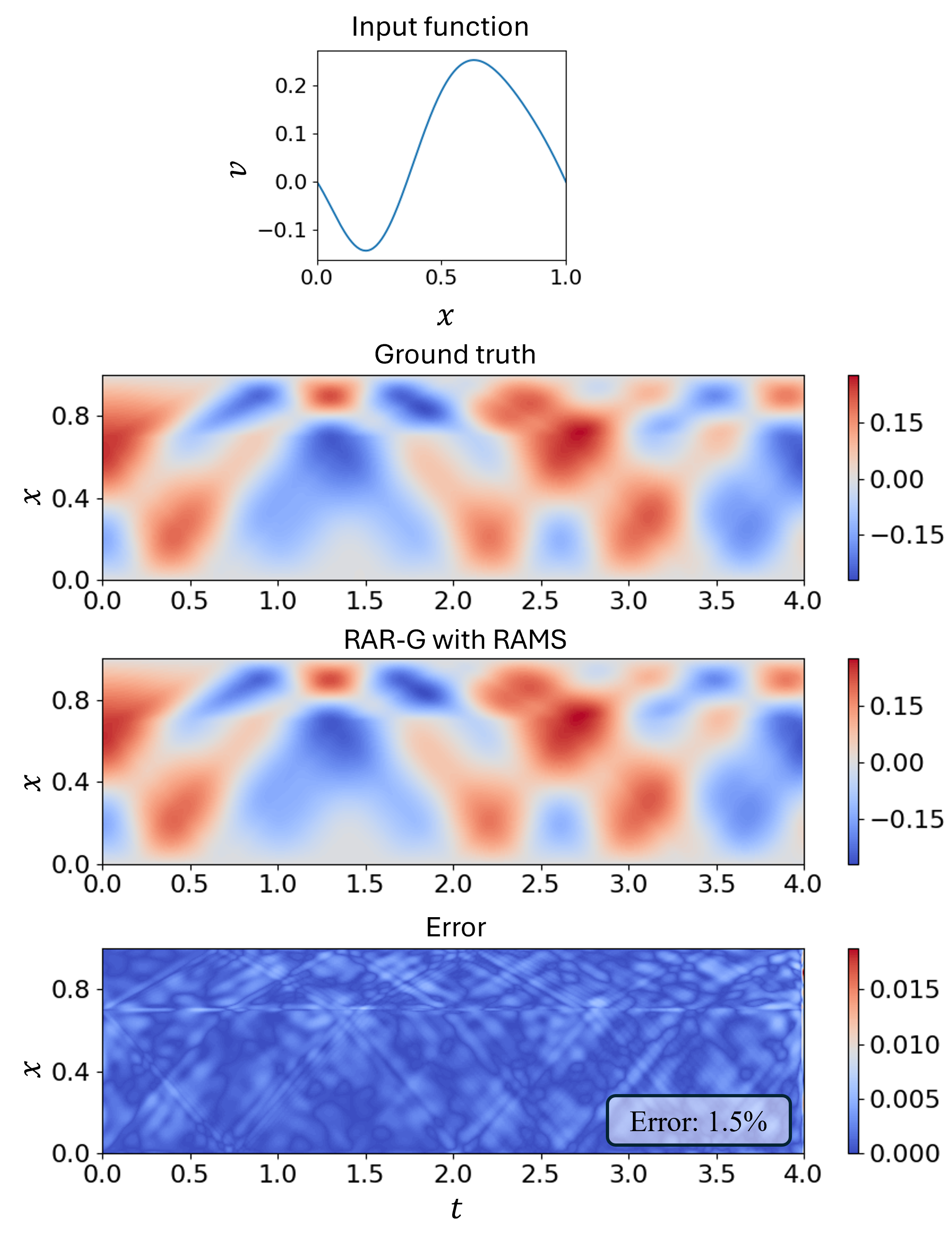}
    \caption{\textbf{Representative examples of the wave equation with discontinuous velocity by data-driven operator learning.}}
    \label{fig:ddol_wave}
\end{figure}

\begin{figure}[htbp]
    \centering
    \includegraphics[width=.9\linewidth]{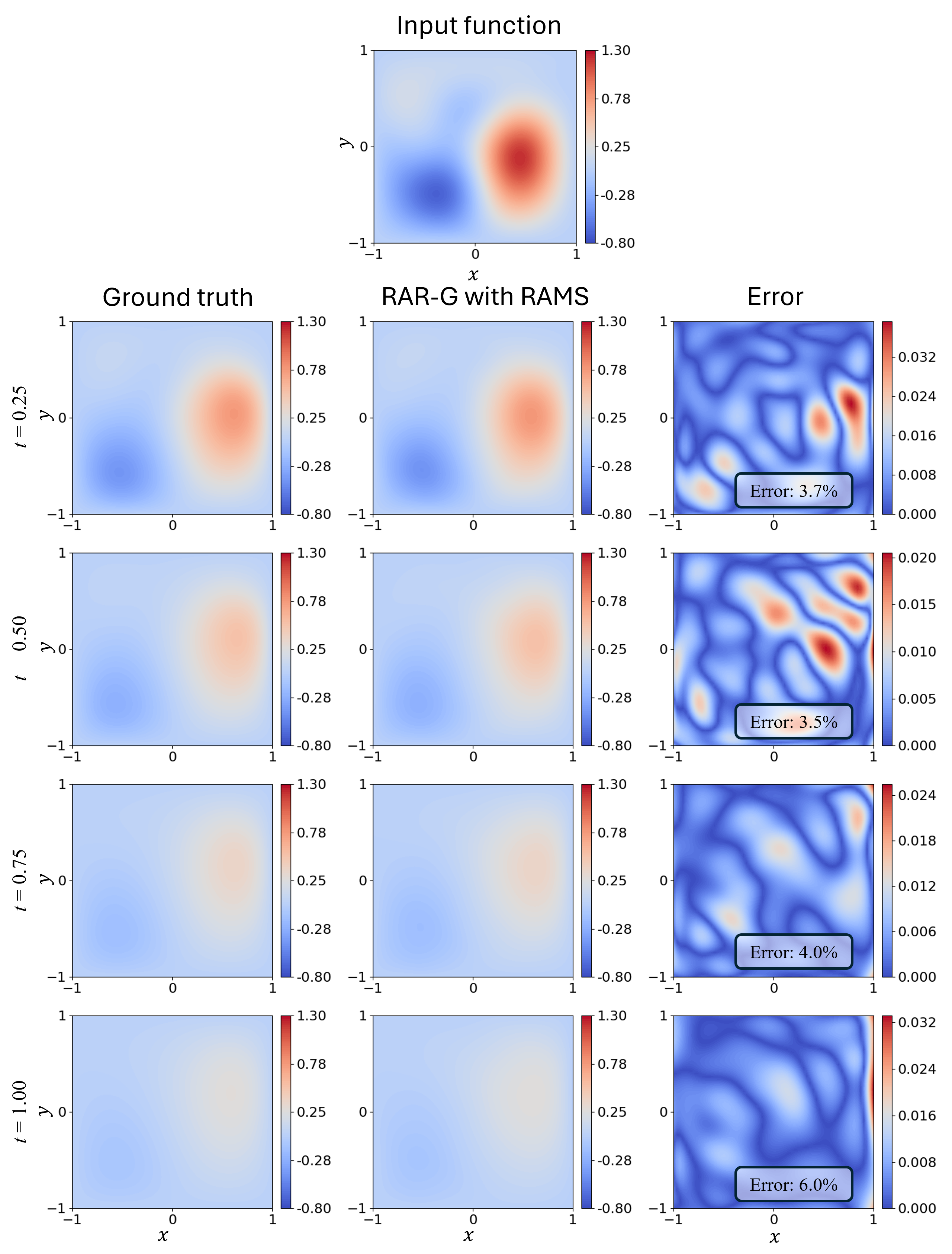}
    \caption{\textbf{Representative examples of two-dimensional Burgers' equation by data-driven operator learning.}}
    \label{fig:ddol_burgers}
\end{figure}

We provide visualizations for our experiments in this section.
\begin{itemize}
    \item Fig.~\ref{fig:pinn_vis}: PINN using RAR-G with RAMS on Burgers' equation (Section~\ref{sec:pinn_burgers}), the wave equation (Section~\ref{sec:pinn_wave}), and the Poisson equation (Section~\ref{sec:pinn_poisson}).
    \item Fig.~\ref{fig:piol_vis}: PI-DeepONet trained with RAMS on the diffusion-reaction equation (Section~\ref{sec:piol_diff}), the advection equation (Section~\ref{sec:piol_adv}; $n_{RAMS}=400$), and the Poisson equation (Section~\ref{sec:piol_poi}; $n_{RAMS}=300$ and $n_{sam}=800$).
    \item Fig.~\ref{fig:ddol_wave}: data-driven DeepONet for the wave equation with discontinuous velocity (Section~\ref{sec:ddol_wave}) by using $n_{RAMS}=400$ and $n_{sam}=200$.
    \item Fig.~\ref{fig:ddol_burgers}: data-driven DeepONet trained using RAR-G with RAMS for the two-dimensional Burgers' equation (Section~\ref{sec:ddol_burgers}; $n_{sam}=3,000$, $t_r=2$, and $p=0.1$ in Case 2).
\end{itemize}

%% file: content/app_abl_proj.tex
\section{Projector in RAMS}
\label{sec:projector_ablation}

\begin{figure}[htbp]
    \centering
    \includegraphics[width=0.7\linewidth]{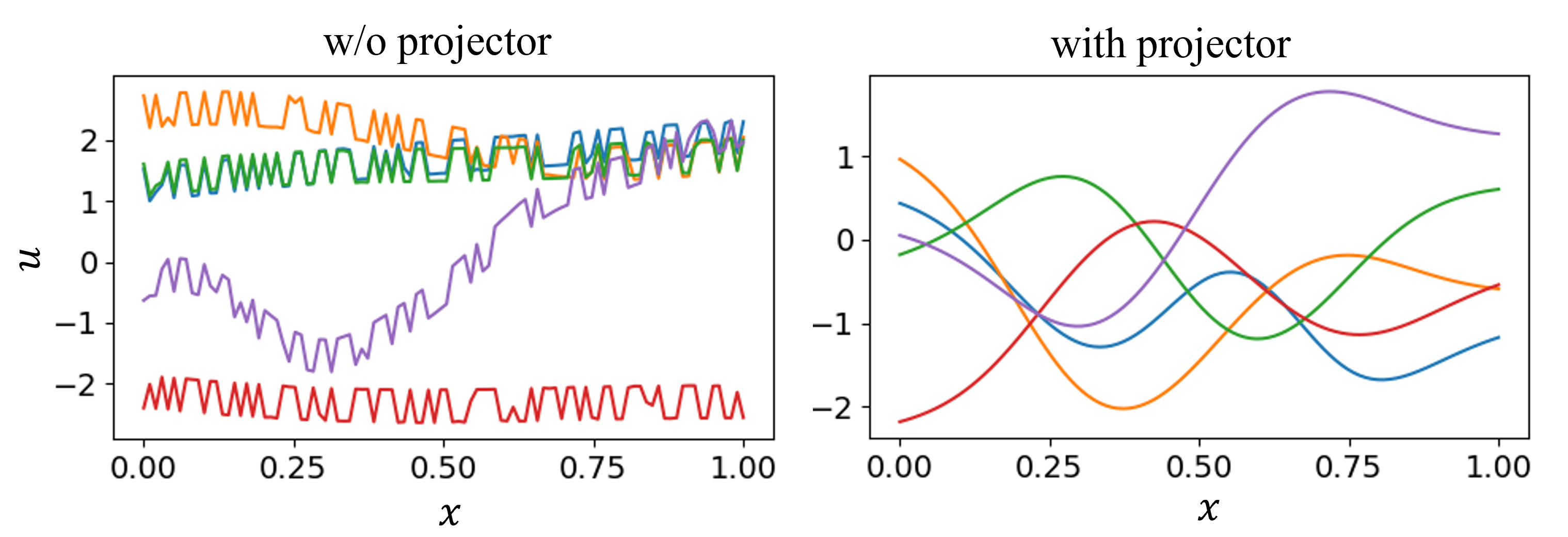}
    \caption{\textbf{Function samples optimized by RAMS with and without projector for the advection equation in PI operator learning.} For each case, five function samples are visualized.}
    \label{fig:projector_ablation}
\end{figure}

We want to demonstrate the necessity of the projector in RAMS for PI operator learning of the diffusion-reaction equation (Section~\ref{sec:piol_diff}). We use the same experimental settings as in Section~\ref{sec:piol_diff}, except for the removal of the function projector component in RAMS. Fig.~\ref{fig:projector_ablation} compares the trained function samples during RAMS with and without the kernel smoothing method as the projector. Samples adjusted without the projector exhibit significant irregular fluctuations, in stark contrast to the stable profiles obtained when using the projector, indicating the importance of the kernel smoothing method in RAMS.

